\documentclass[11pt]{article}
\pdfoutput=1
\usepackage{fullpage}
\usepackage{hyperref}
\usepackage{url}

\usepackage{pdfpages}
\includepdfset{pagecommand={\thispagestyle{plain}}}
\usepackage[title]{appendix}

\newcommand{\Vex}[1]{\vspace{#1ex}}

\newcommand{\mypar}[1]{\Vex{.75}\noindent {\bf #1}}

\title{LPOP: Challenges and Advances in Logic and Practice of Programming}
\author{David S. Warren \hspace{10ex} Yanhong A. Liu\\ 
Computer Science Department, Stony Brook University}
\date{}

\begin{document}

\maketitle

\begin{abstract}
This article describes the work presented at the first Logic and Practice of Programming (LPOP) Workshop, 
which was held in Oxford, UK, on July 18, 2018, in conjunction with the Federated Logic Conference (FLoC) 2018.
Its focus is challenges and advances in logic and practice of programming.
The workshop was organized around a challenge problem that specifies issues in role-based access control (RBAC),
with many participants proposing combined imperative and declarative solutions expressed
in the languages of their choice.

\end{abstract}

\section{Introduction}




The focus of the 2018 Logic and Practice of Programming workshop was on logic and declarative languages for the practice of programming. Of particular interest were languages (1) that have a clear semantic foundation, so that they can be used for concise modeling of complex application problems, facilitating formal proofs and automated analysis, and (2) that are also implementable, so that the implementations can run as specified, as part of real applications. Also of interest were (a) the design of declarative languages, libraries, and tools that facilitate the construction of complex systems and applications, (b) approaches to integrate declarative and procedural programming, and (c) the use of declarative languages to facilitate other programming paradigms, e.g., distributed programming. The target audience for these languages was students who wish to model complex application problems, and practitioners who want to use them.

The goal of the workshop was to bring together the best people and best languages, tools, and ideas to help improve logic languages for the practice of programming and to improve the practice of programming with logic and declarative programming. We prepared to organize the workshop around a number of "challenge problems", including in particular expressing a set of system components and functionalities clearly and precisely using a chosen description language. To that end, we created an extensive challenge for this purpose in the general area of role-based access control.  We also organized invited talks and additional presentations by the proponents of some well-known description methods. We grouped presentations of description methods by the kind of problems that they address, and tried to allow ample time to understand the strengths of the various approaches and how they might be combined.

Potential workshop participants were invited to submit position papers (1 or 2 pages in PDF format), and to state whether they wished to present a talk at the workshop, explaining how they would express the challenge problem. Because we intended to bring together researchers from many parts of logic and declarative languages and practice of programming communities, it was essential that all talks be accessible to non-specialists.

The program committee invited attendees based on their position paper submissions and attempted to accommodate presentation requests in ways that fit with the broader organizational goals outlined above.

\subsection{Program}

The schedule for the presentation of contributed position papers that describe solutions to the challenge problem follows.






\footnotesize
\begin{verbatim}
Session 1: Logic and Practice of Programming
Session Chair: Marc Denecker

09:00 Marc Denecker. Opening and introduction.
09:10 Invited Talk: Michael Leuschel. Practical uses of Logic, Formal Methods, B and ProB.  
09:50 Invited Talk: Nicola Leone, Bernardo Cuteri, Marco Manna, Kristian Reale and Francesco Ricca. 
        On the Development of Industrial Applications with ASP.

Session 2: Security Policies as Challenge Problems
Session Chair: Annie Liu

11:00 Annie Liu. Introduction: Role-Based Access Control as a Programming Challenge. 
11:10 Thom Fruehwirth (in spirit). Discussions on RBAC and 
        "Security Policies in Constraint Handling Rules".
11:20 David S. Warren. LPOP2018 XSB Position Paper.
11:30 Roberta Costabile, Alessio Fiorentino, Nicola Leone, Marco Manna, Kristian Reale 
        and Francesco Ricca. Role-Based Access Control via JASP.
11:40 Marc Denecker. The RBAC challenge in the Knowledge Base Paradigm.
11:50 Tuncay Tekle. Role-Based Access Control via LogicBlox.
12:00 Joost Vennekens. Logic-based Methods for Software Engineers and Business People. 
12:10 Yanhong A. Liu and Scott Stoller. Easier Rules and Constraints for Programming.
12:20 All Workshop Participants. Questions about RBAC challenge solutions.

Session 3: Challenge Solutions and Constraint Solving
Session Chair: K. Tuncay Tekle

14:00 Panel: Practice of Modeling and Programming.
             Panel Chair: Peter Van Roy.  Panelists: All Morning Speakers.              
14:30 Invited Talk: John Hooker. A Modeling Language Based on Semantic Typing.
15:10 Neng-Fa Zhou and Håkan Kjellerstrand. 
        A Picat-based XCSP Solver - from Parsing, Modeling, to SAT Encoding.
15:20 Paul Fodor. Role-Based Access Control as a LP/CP/Prolog Programming Challenge.

Session 4: Logic and Constraints in Applications
Session Chair: David Warren

16:00 Invited Talk: Rustan Leino. The Young Software Engineer’s Guide to Using 
        Formal Methods.
16:40 Torsten Schaub. How to upgrade ASP for true dynamic modelling and solving?
16:50 Peter Van Roy. A software system should be declarative except where it interacts 
        with the real world.
17:00 All Workshop Participants. Questions about logic and constraints in real-world applications.
17:10 Panel: Future of Programming with Logic and Knowledge. 
             Panel Chair: David Warren.  Panelists: All Afternoon Speakers               
17:40 David Warren and Annie Liu. Future of LPOP.
17:50 Tuncay Tekle and Marc Denecker. Closing.
\end{verbatim}
\normalsize

\subsection{Organization}
The organizers and others responsible for the workshop were:\bigskip
\\
\begin{tabular}{@{}l l}
\textbf{Chairs} \\
David Warren &    Stony Brook University \\
Annie Liu &       Stony Brook University \\
\\
\textbf{Program Committee Chairs} \\
Marc Denecker   &   KU Leuven \\
Tuncay Tekle    &   Stony Brook University \\
\\
\textbf{Program Committee} \\
Molham Aref &         Relational AI \\
Manuel Carro &        IMDEA Software \\
Thomas Eiter &        Technical University of Vienna \\
Jacob Feldman  &        OpenRules \\
Thom Frühwirth  &        University of Ulm \\
Michael Kifer   &           Stony Brook University \\
Mark Miller     &            Google \\
Enrico Pontelli  &          New Mexico State University \\
Francesco Ricca  &      University of Calabria \\
Peter Van Roy    &         Université catholique de Louvain \\
Joost Vennekens  &      Katholieke Universiteit Leuven \\
Jan Wielemaker   &      Vrije  Universiteit Amsterdam \\
Neng-Fa Zhou     &       City University of New York \\
\end{tabular}

\subsection{Homepage}
\url{http://lpop.cs.stonybrook.edu/}\\
It contains the full workshop program with links to the presentation slides.

\section{Invited Talks}

Four invited speakers gave excellent talks:\smallskip\\
\begin{tabular}{@{}l l}
John Hooker &     Carnegie Mellon University \\
Rustan Leino &    Amazon Web Services \\
Nicola Leone &    University of Calabria \\
Michael Leuschel &  University of Dusseldorf \\
\end{tabular}

\subsection{A Modeling Language Based on Semantic Typing}
\mypar{Speaker: John Hooker, Carnegie Mellon University}

\mypar{Abstract:}
A growing trend in modeling is the construction of high-level modeling languages that invoke a suite of solvers. This requires automatic reformulation of parts of the problem to suit different solvers, a process that typically introduces many auxiliary variables. We show how semantic typing can manage relationships between variables created by different parts of the problem. These relationships must be revealed to the solvers if efficient solution is to be possible. The key is to view variables as defined by predicates, and declaration of variables as analogous to querying a relational database that instantiates the predicates. The modeling language that results is self-documenting and self-checks for a number of modeling errors.\\
(Joint work with André Ciré and Tallys Yunes.)

\mypar{Slides:}
\url{https://drive.google.com/file/d/1emvbNY9bp3AWn6h4EHI7y3VphaZZ7gYL/}

\subsection{The Young Software Engineer’s Guide to Using Formal Methods}
\mypar{Speaker: Rustan Leino, Amazon Web Services}

\mypar{Abstract:}
If programming was ever a hermit-like activity, those days are in the past. Like other internet-aided social processes, software engineers connect and learn online. Open-source repositories exemplify common coding patterns and best practices, videos and interactive tutorials teach foundations and pass on insight, and online forums invite and answer technical questions. These knowledge-sharing facilities make it easier for engineers to pick up new techniques, coding practices, languages, and libraries. This is good news in a world where software quality is as important as ever, where logic specification can be used to declare intent, and where formal verification tools have become practically feasible.

In this talk, I give one view of the future of software engineering, especially with an eye toward software quality. I will survey some techniques, look at the history of tools, and inspire with some examples of what can be daily routine in the lives of next-generation software engineers.

\mypar{Slides:}\\
\url{https://drive.google.com/file/d/0B9ffoWLQuWUXRTRtRElodFliMW5uaVhYMGQtb1FiLTJXLTJZ/}

\subsection{On the Development of Industrial Applications with ASP}
\mypar{Speaker: Nicola Leone, University of Calabria}

\mypar{Asbtract:}
Answer Set Programming (ASP) is a powerful rule-based language for knowledge representation and reasoning that has been developed in the field of logic programming and nonmonotonic reasoning. After many years of basic research, the ASP technology has become mature for the development of significant real-world applications. In particular, the well-known ASP system DLV has undergone an industrial exploitation by a spin-off company called DLVSYSTEM srl, which has led to its successful usage in a number of industry-level applications. The success of DLV for applications development is due also to its endowment with powerful development tools, supporting researchers and software developers that simplify the integration of ASP in real-world applications which usually require to combine logic-based modules within a complete system featuring user interfaces, services etc. In this talk, we first recall the basics of the ASP language. Then, we overview our advanced development tools, and we report on the recent implementation of some challenging industry-level applications of our system.\\
(Joint work with Bernardo Cuteri, Marco Manna, Francesco Ricca)

\mypar{Slides:}
\url{https://drive.google.com/file/d/1GGWtDzsIVnh43_kLpPjA8h7P1kDUTkYA/}\\
A paper describing this work is included in Appendix~\ref{leone-invited}. 

\subsection{Practical Uses of Logic, Formal Methods, B and ProB}
\mypar{Speaker: Michael Leuschel, University of Dusseldorf}

\mypar{Abstract:}
The B method is quite popular for developing provably correct software for safety critical railway systems, particularly for driverless trains. In recent years, the B method has also been used successfully for data validation (http://www.data-validation.fr). There, the B language has proven to be a compact way to express complex validation rules, and tools such as predicateB, Ovado or ProB can be used to provide high assurance validation engines, where a secondary toolchain validates the result of the primary toolchain.

This talk will give an overview of our experience in using logic-based formal methods in general and B in particular for industrial applications. We will also touch subjects such as training and readability and the implementation of ProB in Prolog. We will examine which features of B make it well suited for, e.g., the railway domain, but also point out some weaknesses and suggestions for future developments. We will also touch upon other formal methods such as Alloy or TLA+, as well as other constraint solving backends for B, not based on Prolog (SAT via Kodkod/Alloy and SMT via Z3 and CVC4).

\mypar{Slides:}
\url{https://drive.google.com/file/d/1Q19wdAQJiXBTRGiiYM_YjkYqyvEaSvqU/}\\
A Jupyter notebook can be found at:\\
\url{https://drive.google.com/file/d/11UNiLAIlHLHTAmMH__d2JEqrm8kAzc6Z/}

\section{The Challenge Problem}

The domain and the specific functions and components of the challenge problem were selected to give participants the opportunity to demonstrate the best features of their (preferred) logic language. Those features may be from a broad spectrum: elegance, naturalness, compactness, modularity of expression, broadness of the functionality of the logic tools (e.g., a strong point would be if tools are available to prove correctness of your solutions), reuse of the specification to solve different parts of the problem, efficiency, etc.

Participants were free to select only a subset of the functions and components, or to implement variants of them, as long as their solutions showed the utility of their logic approach.

The domain of the challenge was Role-Based Access Control (RBAC). This is a security policy framework for controlling user access to resources based on roles. The challenge included functions and components for several well-known variants and extensions of RBAC, each involving its own set of constraints.

Participants were free, indeed encouraged, to present solutions for other components that show specific strengths of their logic, e.g., such as the aforementioned proof of correctness of logic solutions. We were interested as well in new challenges for logic systems, tasks that cannot yet be solved by existing systems but that pose an interesting research goal.

The RBAC programming challenge is included in Appendix~\ref{liu-challenge}.
The slides for the presentation are in the first half of those at:\\
\url{https://drive.google.com/file/d/1kzfE_CTYfAYgGSLg75ZJojhF1fEk4BSC/}

Participants were encouraged to include programs, specifications, and other related materials in appendices to their position papers.
These papers appear as appendices.

\section{Solutions to the Challenge}

We summarize each proposed solution to the RBAC challenge in the following sections. 

\subsection{Answer Set Programming with Java: JASP}

Nicola Leone presents joint work with Roberta Costabile, Alessio Fiorentino, Marco Manna, Kristian Reale, and Francesco Ricca that attacks the RBAC challenge problem using Answer Set Programming (ASP), as implemented in the JASP system.  JASP is an extension of the Java programming language with an ASP solver, allowing a programmer to use Java for procedural, state-changing operations and use ASP for declarative query solving. 
To solve the RBAC challenge problem, Java is used to update an external relational database and to read the state of the database, generate the necessary facts and rules in the correct form required for the ASP solver, invoke the ASP solver to compute query answers declaratively, and finally update the external database based on the query results when necessary.
This approach separates the procedural aspects of the problem from the query aspects by implementing the procedural aspects in the procedural language Java and the query aspects in the declarative ASP framework.

Leone's presentation describes in detail the issues around implementing the RBAC function {\em GetRolesShortestPlan}, which he says is "the hardest function" to implement in this framework.  It is mentioned that other tasks of the challenge can be solved in a similar way but no specifics are given.
The slides for the presentation of Leone, et al. can be found at:\\
\url{https://drive.google.com/file/d/1cBOHB4Vj3QS21iwp_-fLB36-KUv8ZEGF/}

\subsection{Prolog with Tabling: XSB}
David S. Warren approaches the RBAC challenge problem using classical Prolog, in particular, the version implemented in the XSB system \cite{XSB-programmers-manual}, taking advantage of particular features of that implementation.
The traditional approach would be to use Prolog's {\em assert} and {\em retract} operations to update RBAC facts stored in Prolog's global internal database.  
But this approach is non-declarative.
So instead Warren uses a data structure to represent a database state that is explicitly passed through all defined update operations and query operations required to solve the challenge problem.
This makes them all purely declarative.  
The procedural aspects are integrated into the declarative logical framework by making all update predicates depend on input and output database arguments.
This can be seen as a primitive implementation of a kernel portion of Transaction Logic \cite{Bonner1993TransactionLP}.
Integrity constraint checking can be done before or after database update, with Prolog's standard backtracking naturally handling "transaction rollback" if a check fails.
Warren notes that Prolog's DCG notation can be used to avoid having to explicitly pass the database parameter through all update operations.

The main challenge with this approach is that of efficiency, i.e., whether this Prolog data structure can compete in efficiency with Prolog's native {\em assert} and {\em retract} and whether tabling, which is fundamental to XSB's evaluation strategy, can be made efficient when applied to predicates containing the database as argument(s).
These problems are attacked by the use of a data structure defined in a new XSB package that supports update and query operations on a set of Prolog rules stored in a complex, trie-based Prolog term, but no detailed discussion of performance is provided.

Warren's RBAC implementation solved all the update and direct query and aggregation problems proposed in the challenge; he did not attempt the more complex optimization problems.  

Warren's paper and RBAC solutions are included in Appendix~\ref{warren}.
His presentation slides are available at:\\
\url{https://drive.google.com/file/d/1DhgLh4LkUCs3JcrieqcPTdb_hBL8yoO4/}

\subsection{Knowledge Base Paradigm: IDP}
Marc Denecker, in joint work with Jo Deviendt, describes an approach to solving the RBAC challenge problem in the framework of IDP, an implementation of a knowledge base paradigm.
IDP uses first-order logic combined with inductive definitions to specify declarative knowledge, and then applies a variety of inference mechanisms to this static data to solve various knowledge problems.  
The difficult aspect of the RBAC tasks for this framework is how to incorporate the database update operations within this purely logical paradigm.  
This presentation is a theoretical exploration of how this might be done in the IDB framework; no actual code for any RBAC task is provided.

The approach taken here is to add an explicit temporal argument to each predicate that describes the RBAC state.
Thus the procedural aspects of the problem are handled by using a temporal logic and explicitly reasoning with time.
Then the framework needs "boiler-plate" frame axioms that describe the important properties of time, and also axioms that describe the properties of time for predicates that contain a temporal argument.
Finally, one must explore how the axioms can be efficiently processed by the inference mechanisms of IDP to ensure that this approach will lead to practical solutions to the various tasks of the RBAC challenge.

The paper is provided in Appendix~\ref{denecker}.
The slides of Denecker's presentation are available at: \\
\url{https://drive.google.com/open?id=1Q5JHPuAwWPhBIBdbO_JyUikIeDlMswfy/}

\subsection{Datalog Extensions and Scripting Blocks: LogicBlox}

Tuncay Tekle presented a solution to the RBAC challenge using LogicBlox, a
commercial system for developing enterprise transactions and analytics
applications.  The solution was enabled by LogicBlox's powerful query language,
LogiQL, which extends Datalog with constraint checking, aggregates, and
updates.  

Tekle summarizes LogicBlox as "a state-based system with a persistent
database that can be manipulated, where one can add facts to the database,
and rules and constraints to the state, and query the database at any point
in time".  LogicBlox also uses command line scripting to execute blocks of rules, facts, etc.

For the RBAC challenge, specifications of sets and relations, and
constraints over them are easily written in LogiQL.  So are relational
queries over them, including recursive queries, all expressed easily using
Datalog rules.  Aggregations such as count are expressed use special, extended forms
of rules, less direct than can be expressed using SQL.  Updates are
directly expressed with notations + and - in the conclusions of rules.

For the two optimization problems in administrative RBAC, a restricted
version of one of them could be expressed in the LogicBlox framework with
some rewrite.  The other optimization problem and the two planning problems
could not be solved using LogicBlox.

Tekle's paper is included in Appendix~\ref{tekle}.
The slides are available at:\\
\url{https://drive.google.com/open?id=1sp5poNjknmNVbkNhHYhyvupPY9VTeGyn/}

\subsection{Logic with Interface: IDP with Python API, or DMN}

Joost Vennekens illustrated solving the RBAC challenge using an approach
he had recently proposed.
In this approach, a relation is represented as a list of tuples, directly
written as so in the Python programming language, and a relational query is expressed using a Python
generator expression, such as "all" for universal quantification.  This way,
programmers need to know only the programming constructs in Python, not those in
logic programming systems.

These programming constructs are taken as an interface to a logic
programming systems, where the data and queries could be interpreted
with a more general meaning, e.g., as constraints relating the data,
instead of queries of some derived data from given data.  This general
meaning allows some desired derived data be given and some other data be
inferred.

Vennekens had developed such a Python API for the IDP system.  He
expresses in Python two example relations and an example query from core RBAC in the RBAC
challenge, but not the rest of the functions and components.

Vennekens also uses the recent Decision Model and Notation (DMN) standard
to support the general argument that more familiar notations to domain
experts can help increase the impact of logic-based methods to business people.

This paper is provided in Appendix~\ref{vennekens}. 
The slides can be found at:\\
\url{https://drive.google.com/open?id=1WYQjkfS1blOU5zvM5ZAzDm12akSGAoBk/}

\subsection{Rules and Constraints Extending Python: DistAlgo Extensions}

Annie Liu presented in joined work with Scott Stoller a solution to the RBAC
challenge in a high-level language that extends the Python programming
language.  This work starts with DistAlgo, an extension of Python for
distributed programming especially with high-level set and logical queries,
and proposes to add rules, constraint optimizations, and backtracking.

With DistAlgo, their solution specifies the hierarchical component
structure of the challenge RBAC explicitly, as required in the challenge
and as in the ANSI standard.  This includes core RBAC, hierarchical RBAC,
core RBAC with constraints, hierarchical RBAC with constraints, and
Administrative RBAC, as in the main challenge, as well as distributed RBAC
as an optional component in the challenge.

Each component includes the definitions of sets and relations, in addition
to those inherited from the parent components if any, as well as all query
and update operations.  For computing transitive closures in hierarchical
RBAC, they gave an implementation that uses high-level set queries and an
alternative implementation that uses Datalog rules.

All components and operations are fully executable in DistAlgo except for
administrative RBAC, which needs the extensions for constraint optimization
and backtracking, and the alternative implementation of transitive closure
using rules.

This paper including the solution program is provided in Appendix~\ref{liu}. 
The slides for the presentation are in the second half of those at:\\
\url{https://drive.google.com/file/d/1kzfE_CTYfAYgGSLg75ZJojhF1fEk4BSC/}

\subsection{RBAC Role Minimization as a LP/CP Programming Contest Challenge}

Paul Fodor presented different logic programming solutions to the
problem of minimum role assignments with hierarchy in
Administrative RBAC.  This is formulated as a constraint optimization problem, and as such
does not address the issues of state update and imperative programming.
Fodor used this problem as the first problem in the Logic
Programming and Constraint Programming Contest at ICLP
2018 (\url{https://sites.google.com/site/prologcontest2018/}).
He presented the best four solutions, two in ASP, one in Prolog, and one in Picat.
The two ASP solutions both used the \#minimize operator, but defined the predicate to 
be optimized in somewhat different ways. 
The Prolog solution used tabling to find all possible solutions, findall to collect them, and then 
explicit comparisons to find the optimal one.
The Picat solution formulated the problem as a constraint problem, 
similar in concept to the ASP solutions, but using Picat's syntax and primitives.

This does not have a paper. The slides can be found at:\\
\url{https://drive.google.com/file/d/1aszplEMEUdUyaUqNU8_GhFtzbLN42dmf/}

\subsection{Security Policies in Constraint Handling Rules (CHR)}

Thom Fruhwirth provided a position paper on the use of CHR 
for the representation of security policies, but he did not provide a solution to the RBAC challenge.  He was unable to attend the workshop to give a presentation.
His paper can be found in Appendix~\ref{fruhwirth}.

\section{Additional presentations}

Some authors and presenters did not address the RBAC challenge but discussed methods, tools, and ideas for integrating different programming paradigms.  We summary each of these below.

\subsection{Upgrading ASP for True Dynamic Modelling and Solving}

In his presentation "How to upgrade ASP for true dynamic modelling and solving" Torsten Schaub discusses the issues involved in extending ASP concepts and implementations in ways that support the solving of dynamic problems, i.e., problems that involve data that change over time.  
He discusses three important aspects of extending ASP in this direction: modeling, encoding and solving, and bench-marking.
Modeling issues involve what formal extension to the logic of ASP is appropriate for specifying dynamic systems.  
Schaub proposes Temporal Equilibrium Logic, which combines the ideas of the logic of Here-and-There with Linear Temporal Logic.
Encoding involves how to represent a problem in the modeling language in such a way that it can be efficiently solved by ASP solvers and their extensions.  
And finally Schaub emphasizes the importance of good, scalable, realistic benchmarks that allow various systems to be effectively compared.
He proposes that benchmarks be developed to address the real-world problem of controlling warehouse operations that use robot vehicles to retrieve items from mobile shelves.  
He argues that this provides an excellent domain for exploring many aspects of using an ASP framework for dynamic systems.

This paper is available in Appendix~\ref{schaub}.
The slides for this talk can be found at:\\
\url{https://drive.google.com/file/d/1GUQC4qXYkt9lK3te0Uc6ZrEFgYNDLQuX/view}

\subsection{A Picat-Based XCSP Solver}

Neng-Fa Zhou presented joint work with Hakan Kjellerstrand in a presentation titled "A Picat-based XCSP Solver - from Parsing, Modeling, to SAT Encoding."
The presentation provides an overview of a Picat-based XCSP3 solver, named PicatSAT, which demonstrates the strengths of Picat, a logic-based language, in parsing, modeling, and encoding constraints into SAT.
XCSP3 is an XML-based language for specifying constraint satisfaction problems, and PicatSAT uses Picat to process these specifications.
The presentation included a brief description of parsing the XCSP3 language, the advantages of using specialized Picat constructs to compactly implement a variety of constraints, and issues involved in encoding SAT problems in Picat.

This paper is provided as Appendix~\ref{zhou}.
The slides are available at:\\
\url{https://drive.google.com/file/d/1-F0RwPQVISeqzR1yr1_wkdvGcsT_i4Hq/}

\subsection{Declarative Programming for All Except Interaction with the Real World}

Peter Van Roy proposes a principle for combining declarative programming
and imperative programming when building software systems.
While declarative programming supports ease of reasoning for
analysis, verification, optimization, and maintenance, it cannot
express interaction with the real world, because it does not
support common real-world concepts such as physical time and
named state, which are supported by imperative programming.
Therefore, the principle is: a software system should be
declarative except where it interacts with the real world.

Examples such as the client-server model from distributed
computing are used as motivation, and a formal argument is
outlined using lambda-calculus and an extension.

Van Roy's paper is provided as Appendix~\ref{vanroy}.
The slides for the talk can be viewed at:\\
\url{https://drive.google.com/file/d/1qVrRwsO3b9LJv8OdhCF_Ali98Gpz443t/}

\section{Conclusion}

The workshop was deemed a success, with the panel discussions and audience participation that followed invited talks and paper presentations being particular noteworthy.
The intention of the organizers is to hold LPOP every two years.
LPOP 2020 was initially intended to be held in conjunction with LICS 2020 in Beijing, but due to travel complexities
will instead be held in conjunction with SPLASH 2020.


\bibliography{refs}
\bibliographystyle{plain}

\begin{appendices}



\section*{Supplementary material}

# Appendix B

# Role-Based Access Control as a Programming Challenge

Yanhong A. Liu*

This programming challenge description focuses on a small but rich set of problems from an important practical application domain, Role-Based Access Control (RBAC). The goal is to allow the use of a wide variety of essential programming constructs to first specify the problems clearly and then solve the problems efficiently, as much as possible.

- Role-Based Access Control (RBAC) is a security policy framework for controlling user access to resources based on roles [3, 9]. It is extremely important for reducing the cost of policy administration, especially in large organizations.

- The problems include updates, for actions and transactions, and queries, for checking, analysis, optimization, and planning, in the presence of constraints, naturally organized into a set of components for ease of use by the applications.

The RBAC programming challenge is described in the next two pages.

Among the five RBAC components described, functionalities of the first four are created based on the ANSI standard for RBAC [4, 1] but reduced to contain only the most essential concepts and improved to avoid discovered anomalies [8, 6]. Functionalities in the last component are created to correspond to role mining [2] and generalize from user-role reachability [10].

- As a programming challenge, any subset of self-contained components and functionalities can be used, and the rest can be made optional.

- Additional RBAC components and functionalities can also be added, for example, for sessions and for dynamic separation of duty (DSD) constraints in the ANSI standard [4, 1], for role mining with probabilistic models [5], and for trust management [7] (also called distributed RBAC) in decentralized systems.

- Furthermore, one may add a verification component for proving or checking the constraints, a Graphical User Interface (GUI) component, a particular RBAC policy for an RBAC system, and a test component for correctness and performance testing.

This programming challenge is created for the Workshop on Logic and Practice of Programming (LPOP) at the Federated Logic Conference (FLOC), Oxford, UK, July 18, 2018. The emphasis is on clearly expressing the problem logic first before improving the program efficiency. Any languages and systems can be used.

---
*Author's contact: Computer Science Department, Stony Brook University, Stony Brook, New York. Email: liu@cs.stonybrook.edu

# RBAC programming challenge

We consider Role-Based Access Control (RBAC) with 5 components:
   Core RBAC
   Hierarchical RBAC
   Core RBAC with Static Separation of Duty (SSD) constraint (a.k.a. Constrained RBAC)
   Hierarchical RBAC with SSD constraint
   Administrative RBAC

**Core RBAC** keeps several sets including the following:
   `USERS`: set of users
   `ROLES`: set of roles
   `PERMS`: set of permissions
   `UR`: set of user-role pairs
   `PR`: set of permission-role pairs
with constraints:
   `UR` is subset of `USERS * ROLES`
   `PR` is subset of `PERMS * ROLES`
update functions for each set, subject to the constraints above:
   `AddUser`, `DeleteUser`, `AddRole`, `DeleteRole`, `AddPerm`, `DeletePerm`
   `AddUR`, `DeleteUR`, `AddPR`, `DeletePR`, where
   each `Add` has pre-conditions: the element is not in and no constraints will be violated, and
   each `Delete` has the pre-condition that the element is in, and maintains the constraints by
      updates if needed
and query functions including the following:
   `AssignedRoles(user)`: the set of roles assigned to `user` in `UR`
   `UserPermissions(user)`: the set of permissions assigned to the roles assigned to `user`
   `CheckAccess(user, perm)`: whether some role is assigned to `user` and is granted `perm`

**Hierarchical RBAC** extends CoreRBAC and keeps also a role hierarchy:
   `RH`: set of pairs of roles, called ascendant and descendant roles,
         where an ascendant role inherits permissions from a descendant role
with constraints:
   `RH` is subset of `ROLES * ROLES`, and `RH` is acyclic
update functions for `RH`, subject to the constraints above:
   `AddInheritance(asc, desc)`, `DeleteInheritance(asc, desc)`, where
   each update has the same kinds of pre-conditions as updates in CoreRBAC
and query functions including the following:
   `Trans()`: the transitive closure of role hierarchy unioned with the reflexive role pairs
   `AuthorizedRoles(user)`: the set of roles of `user` and their transitive descendant roles

**Core RBAC with SSD** extends CoreRBAC and keeps also a set of SSD items, where
  each item has: a name, a set of roles, and a cardinality
with constraints:
  all roles in all SSD items are in `ROLES`
  for each SSD item, its cardinality is greater than 0 and less than the number of its roles
  for each user, for each SSD item, the number of assigned roles (`AssignedRoles`) of the user
    that are in the item's set of roles is at most the item's cardinality
update functions, subject to the constraints above:
  `CreateSsdSet(name, roles, c)`: add SSD item having `name`, `roles`, and cardinality c
  `DeleteSsdSet(name)`: delete SSD item having `name`
  `AddSsdRoleMember(name, role)`: add `role` to roles of SSD item having `name`
  `DeleteSsdRoleMember(name, role)`: delete `role` from roles of SSD item having `name`
  `SetSsdSetCardinality(name, c)`: set c to be cardinality of SSD item having `name`, where
  each update has the same kinds of pre-conditions as updates in CoreRBAC, except that
    all updates have also pre-conditions that no constraints will be violated
and query functions including the following:
  `SsdRoleSets()`: the set of names of SSD items
  `SsdRoleSetRoles(name)`: the set of roles in SSD item having `name`
  `SsdRoleSetCardinality(name)`: the cardinality of SSD item having `name`

**Hierarchical RBAC with SSD** extends both Hierarchical RBAC and Core RBAC with SSD and combines all from both except that the SSD constraint uses `AuthorizedRoles` in place of `AssignedRoles`

**Administrative RBAC** could extend each of the previous 4 components; we consider extending the last, HierarchicalRBACwithSSD, with optimization and planning functions:
  `MinRoleAssignments`:
    find `ROLES'`, `UR'`, and `PR'` with the smallest total size of `UR'` and `PR'`
    such that each user has the same permission through `AuthorizedRoles` as before
  `MinRoleAssignmentsWithHierarchy`:
    find `ROLES'` , `UR'`, `PR'`, and `RH'` with the smallest total size of `UR'`, `PR'`, and `RH'`
    such that each user has the same permissions through `AuthorizedRoles` as before
  `GetRolesPlan(user, roles, acts)`:
    find a sequence of actions, i.e., updates, in `acts` that allows `user` to get `roles`
  `GetRolesShortestPlan(user, roles, acts)`:
    find a shortest sequence of actions, i.e., updates, in `acts` that allows `user` to get `roles`
and an operation:
  `GetRoles(user, roles, acts)`:
    perform a sequence of actions in `acts` that allows `user` to get `roles` if possible
Any subset of updates can be used as `acts`. All constraints must hold after each update.

## Acknowledgment

# Appendix C

# LPOP2018 XSB Position Paper

David S. Warren
Department of Computer Science
Stony Brook University, New York, USA

May 30, 2018

## 1 Introduction

In this position paper we first describe a classic logic programming approach to the solution of (a portion) of the challenge problem involving RBAC. We use the XSB Tabled Prolog Language and system [3], with ideas from Transaction Logic [1]. Then we discuss efficiency and scalability issues for this implementation. Finally we discuss issues that involve what would be required to use such an implementation in a real-world application requiring RBAC functionality.

## 2 RBAC Challenge Problem in XSB

We describe our solution to the challenge problem. We use a module, `prolog_db`, that was recently added to the XSB system that allows a Prolog database (i.e., a set of clauses) to be represented as a ground term, which we call a PDB. A number of operations are provided to access and update PDB's, the salient ones here being a) `assert_in_db(+Clause,+PDB0,-PDB)`, which adds a clause to a PDB to generate a new PDB, b) `retract_in_db(+Clause,+PDB0,-PDB)`, which deletes a clause from a PDB to generate a new PDB, and c) `call_in_db(?Goal,+PDB)`, which calls a goal in a given PDB, returning instances that are true in the given PDB. For this RBAC application the clauses in a PDB will always be ground facts. We use the Prolog Definite Clause Grammar (DCG) notation for writing these programs, since it supports a convenient notation for writing rules that define state transformations. The (implicit) state is always a PDB.

The description of the RBAC challenge problem is given at
https://drive.google.com/file/d/1q9W15kI624TI6pEbh2IMPDw6X5_5MiW7/view.
The XSB program for the RBAC challenge problem (minus the two `MinRoleAssignment` functions in the Administrative component) is provided in the Appendix.

This is a relatively straightforward specification (and implementation) of the problem in classical logic programming. Since the RBAC database is represented explicitly as a term in Prolog, general Prolog backtracking restores earlier database states. So this makes post conditions, such as in `create_ssdSet`, trivial to implement: do the operation, and then check the post condition; if it fails, the system automatically backtracks to restore the initial database state.

Note also that the tabling is correct even as the PDBs change, since the appropriate PDBs, which are implicit in the DCG notation, are arguments to the tabled predicates. One might want to abolish the tables periodically if space becomes an problem.

## 2.1 Performance Issues

This implementation should be quite efficient as an XSB program. A PDB represents a set of clauses. The `prolog_db` module uses a trie data structure to store a PDB, with a variant of a radix tree at each branch point in the trie. This makes the representation canonical, in that a given set of clauses is represented by the same term, regardless of the sequence of asserts and retracts (`_in_db`) that generates that set. So all updates and accesses are done in log time. Also, the terms representing PDBs are ground and so can use "interned terms", also sometimes known as hash-consing, which are implemented in XSB [2]. Thus the terms are copied to a global store and uniquely stored; i.e. all common subterms are shared. Then the Prolog code passes around what are essentially pointers to tries in the global store. With this representation tabling involves only the constant-time copying of a "pointer" into and out of a table. Also equality comparison of two PDBs is simply a comparison of their pointers. The `GetRoles(Shortest)Plan` functions do an exhaustive search for plans, which in some cases could be expensive, but the tabling does provide help. The two `MinRoleAssignments` functions are not implemented because they seem to require constraint solving, which is not XSB's strength. An exhaustive search could be implemented directly in XSB but would be uninteresting.

## 2.2 Interfacing with the System Environment

So this seems to us to be a reasonably elegant solution to the formal RBAC problem as specified in (all but two functions of) the challenge. However, there is the question of how this code might really be used in a much larger system in which access control is only a small component. As described in the previous subsection, we don't think that the performance and scalability of the execution of the RBAC operations would present a problem. The more difficult issues, we believe, involve data persistence and concurrent usage. There are various potential solutions, but no single obvious one (at least to us.) And the potential solution seem to require procedural, more than logical, thinking and programming.

# A  RBAC Implementation in XSB

```prolog
:- import assert_in_db/3, retractall_in_db/3, call_in_db/2, size_db/2, new_dbi/1
    from prolog_db.

%% rename update operations for clarity (and brevity)
add(Fact,DB0,DB) :- assert_in_db(Fact,DB0,DB).
del(Fact,DB0,DB) :- retractall_in_db(Fact,DB0,DB).

%% CORE RBAC
%% Make relation lookups into identity transactions (convenience)
%%   i.e., they return the exact database they receive.
users(User,D,D) :- call_in_db(users(User),D).
roles(Role,D,D) :- call_in_db(roles(Role),D).
perms(Perm,D,D) :- call_in_db(perms(Perm),D).
ur(User,Role,D,D) :- call_in_db(ur(User,Role),D).
pr(Perm,Role,D,D) :- call_in_db(pr(Perm,Role),D).

%% update functions
%% to add and delete users...
addUser(User) --> \+ users(User), add(users(User)).
deleteUser(User) --> users(User), \+ ur(User,_), del(users(User)).

%% to add and delete roles...
addRole(Role) --> \+ roles(Role), add(roles(Role)).
deleteRole(Role) --> roles(Role), \+ ur(_,Role), \+ pr(_,Role), del(roles(Role)).

%% to add and delete permissions...
addPerm(Perm) --> \+ perms(Perm), add(perms(Perm)).
deletePerm(Perm) --> perms(Perm), \+ pr(Perm,_), del(perms(Perm)).

%% to add and delete users in roles
addUR(User,Role) --> users(User), roles(Role), \+ ur(User,Role), add(ur(User,Role)).
deleteUR(User,Role) --> ur(User,Role), del(ur(User,Role)).

%% to add and delete permissions that roles have.
addPR(Perm,Role) --> perms(Perm), roles(Role), \+ pr(Perm,Role), add(pr(Perm,Role)).
deletePR(Perm,Role) --> pr(Perm,Role), del(pr(Perm,Role)).

%% simple rename as required
assignedRoles(User,Role) --> ur(User,Role).

%% define user permissions by joining user-role and role-permission
userPermissions(User,Perm) --> ur(User,Role), pr(Perm,Role).

%%%%%%%%%%%%%%%%%%%%%%%%%%%%%%%%%%%%%%%%%%%%%%
%% HIERARCHICAL RBAC (additions)

%% immediate subclass relation to identity transaction
%% (would be more efficient in other order..)
rh(RoleAsc,RoleDsc,D,D) :- call_in_db(rh(RoleAsc,RoleDsc),D).

%% update functions
%% add an inheritance fact if no loop is generated
addInheritance(RoleAsc,RoleDsc) -->
    roles(RoleAsc),roles(RoleDsc),
    add(rh(RoleAsc,RoleDsc)),
```

```
    \+ trans(RoleAsc,RoleDsc).

%% define transitive closure for inheritance, and loop checking
%% note that the db is a (hidden) parameter, so this is correct over updates
:- table trans/4.
trans(Dsc,Dsc) --> [].
trans(Dsc,Asc) --> trans(Dsc,Par),rh(Asc,Par).

%% remove an inheritance fact.
deleteInheritance(RoleAsc,RoleDsc) -->
    rh(RoleAsc,RoleDsc),
    del(rh(RoleAsc,RoleDsc)).

%% add rule to include inheritance when determining authorized roles
authorizedRoles(User,Role) -->
    trans(Role,ARole),
    assignedRoles(User,ARole).

%%%%%%%%%%%%%%%%%%%%%%%%%%%%%%%%%%%%%%%%%%%%
%% SSD

%% again make lookup operations into identity transactions, for convenience.
ssdCount(Name,Cnt,D,D) :- call_in_db(ssdCount(Name,Cnt),D).
ssdRole(Name,Role,D,D) :- call_in_db(ssdRole(Name,Role),D).

%% add ssdRole and check consistency
addSsdRoleMember(Name,Role) -->
    roles(Role),
    add(ssdRole(Name,Role)),
    ssdConsistent(Name).

%% delete ssdRole
deleteSsdRoleMember(Name,Role) -->
    ssdRole(Name,Role),
    del(ssdRole(Name,Role)).

%% set SSD cardinality
setSsdSetCardinality(Name,Cnt) -->
    (ssdCount(Name,OCnt)              % if already has a cardinality
     ->({OCnt \== Cnt}                 % if cardinality is changed
         -> del(ssdCount(Name,OCnt)),
            add(ssdCount(Name,Cnt)),
            ssdConsistent(Name)     % check consistency after update
        ; []                          % no change, no need to check consistency
       )
     ; add(ssdCount(Name,Cnt)),     % create initial cardinality
ssdConsistent(Name)         % check consistency
    ).

%% ssd set ops
deleteSsdSet(Name) -->
    ssdCount(Name,_),
    del(ssdCount(Name,_)),
    del(ssdRole(Name,_)).

create_ssdSet(Name,RoleList,Cnt) -->
    \+ ssdCount(Name,_),
```

```
    add(ssdCount(Name,Cnt)),
    addSsdRoleMembers(Name,RoleList),
    ssdConsistent(Name).

%% add role members
addSsdRoleMembers(_,[]).
addSsdRoleMembers(Name,[Role|Roles]) :-
    roles(Role),
    add(ssdRole(Name,Role)),
    addSsdRoleMembers(Name,Roles).

%% define consistency.
ssdConsistent(Name) --> ssdConsistentAssigned(Name).

%% or if include authorized, would use the following clause:
%% ssdConsistent(Name) --> ssdConsistentAuthorized(Name).

%% consistent if not inconsistent
ssdConsistentAssigned(Name) -->
    \+ ssdInconsistentAssigned(Name).

%% inconsistency is more easily defined
ssdInconsistentAssigned(Name) -->
    ssdCount(Name,Cnt),
    ssdAssignedRoleCnt(Name,_User,UCnt),
    {UCnt > Cnt}.

%% aggregation predicate for counting roles
sum(A,B,C) :- C is A + B.

%% aggregate using sum (admittedly, a bit awkward...)
:- table ssdAssignedRoleCnt(_,_,fold(sum/3,0),_,_).
ssdAssignedRoleCnt(Name,User,1) -->
    assignedRoles(User,Role),
    ssdRole(Name,Role).

%% same as above but using authorized as opposed to assigned.
ssdConsistentAuthorized(Name) -->
    \+ ssdInconsistentAuthorized(Name).

ssdInconsistentAuthorized(Name) -->
    ssdCount(Name,Cnt),
    ssdAuthorizedRoleCnt(Name,_User,UCnt),
    {UCnt > Cnt}.

:- table ssdAuthorizedRoleCnt(_,_,fold(sum/3,0),_,_).
ssdAuthorizedRoleCnt(Name,User,1) -->
    authorizedRoles(User,Role),
    ssdRole(Name,Role).
```

```
%%%%%%%%%%%%%%%%%%%%%%%%%%%%%%%%%%%%%%%%%%
%% Administrative RBAC (Planning functions)

%% Exhaustive search for a plan
:- table getRolesPlan/6.
getRolesPlan(User,Roles,Acts,PlanSet) -->
    (hasAllRoles(User,Roles)    % at the goal, so
     ->{new_dbi(PlanSet)}       % no actions needed
     ; {member(Act,Acts)},      % for each action
        call(Act),              % perform it
        getRolesPlan(User,Roles,Acts,PlanSet0),  % search from new DB
        {assert_in_db(Act,PlanSet0,PlanSet)}  % succeeded, add this act
    ).

%% check goal state
hasAllRoles(_,[]) --> [].
hasAllRoles(User,[Role|Roles]) -->
    authorizedRoles(User,Role),
    hasAllRoles(User,Roles).

%% Search for shortest plan:
%% Same as above, but only keep plans with fewest actions
:- table getRolesShortestPlan(_,_,_,lattice(smaller_plan/3),_,_).
getRolesShortestPlan(User,Roles,Acts,PlanSet) -->
    (hasAllRoles(User,Roles)
     ->{new_dbi(PlanSet)}
     ; {member(Act,Acts)},
        call(Act),
        getRolesShortestPlan(User,Roles,Acts,PlanSet0),
        {assert_in_db(Act,PlanSet0,PlanSet)}
    ).

%% Plan2 is the smaller of Plan0 and Plan1
smaller_plan(Plan0,Plan1,Plan2) :-
    size_db(Plan0,N0),
    size_db(Plan1,N1),
    (N0 =< N1
     ->Plan2 = Plan0
     ; Plan2 = Plan1
    ).
```

# Appendix D
# Role-Based Access Control via JASP

Roberta Costabile[1], Alessio Fiorentino[1], Nicola Leone[1,2],
Marco Manna[1], Kristian Reale[1,2], and Francesco Ricca[1]

[1]University of Calabria, Italy – `<lastname>@mat.unical.it`
[2]DLVSystem s.r.l, Italy

**Abstract.** In this paper, we answer the Role-Based Access Control (RBAC) challenge by showcasing the solution of RBAC components by using JASP, a flexible framework integrating ASP with Java. In JASP the programmer can simply embed ASP code in a Java program without caring about the interaction with the underlying ASP system. This way, it is possible solve seamlessly both tasks suitable for imperative and declarative specification as required by RBAC.

## 1 Introducing JASP

Answer Set Programming (ASP) [1] is a fully-declarative logic programming paradigm proposed in the area of knowledge representation and non-monotonic reasoning. Its idea is to represent a given computational problem by a logic program whose answer sets correspond to solutions, and use a solver to find them. After many years of research, the formal properties of ASP are well-understood; notably, ASP is expressive: it can solve problems of complexity beyond NP [4, 5]. Moreover, the availability of robust and efficient solvers [8] made ASP an effective powerful tool for advanced applications, and stimulated the development of many interesting applications [6]. Recently, we have employed ASP for developing some industrial application, such as: building systems for workforce management [10], e-tourism [3], and solving complex industry-relevant problems [2].

The experience we gained has confirmed the viability of the industrial exploitation of ASP. However, it has evidenced the strong need of integrating ASP technologies (i.e., ASP programs and solvers) with well-assessed software-development processes and platforms, which are tailored for modern imperative object-oriented programming languages [9]. Indeed, the lesson we have learned, while building real-world ASP-based applications, confirms that complex business-logic features can be developed in ASP at a lower (implementation) price than in traditional imperative languages. Indeed, from a software engineering viewpoint, the employment of ASP brings many advantages not only in terms of readability, but also in flexibility, extensibility, and ease of maintenance. However, since ASP is not a fully general-purpose language, declarative specifications have to be "embedded", at some point, inside imperative modules that are necessary to develop user-friendly applications making use, for example, of visual user-interfaces. To this end, we have introduced a new programming framework integrating ASP with Java [7]. The framework is based on a hybrid language,

called JASP, that transparently supports a bilateral interaction between ASP
and Java. The programmer can simply embed ASP code in a Java program without caring about the interaction with the underlying ASP system. The logical
ASP program can access Java variables, and the answer sets, resulting from the
execution of the ASP code, are automatically stored in Java objects, possibly
populating Java collections, transparently.

## 2 Modeling the RBAC Challenge with JASP

In this section, we sketch a JASP-based solution to the Role-Based Access Control (RBAC) challenge.[1] According to JASP's philosophy, we associate a Java
class to each RBAC-set occurring in the five components. Then, we define a class
Manager to implement all methods performing updates and queries from checking to planning. In what follows, we report the implementation of method trans
—computing the transitive closure of role hierarchy (encoded via relation rh)
unioned with the reflexive role (encoded via relation role) pairs— to appreciate
the succinctness and elegance of the approach:

```
public List<RH> trans() {
    List<RH> jtr = new ArrayList<RH>();
    <#
       IN = jrole::role;
       IN = jrh::rh;
       OUT = jtr::tr;
       tr(R, R) :- role(R).
       tr(R1, R2) :- rh(R1, R2).
       tr(R1, R3) :- tr(R1, R2), tr(R2, R3).
    #>
    return jtr;
}
```

In particular, we create the Java object jtr and state, via the keyword OUT, that
it will host —after the subsequent three ASP rules will have been evaluated—
all tuples of predicate tr, which occurs in the head of those ASP rules. Then, we
use the keyword IN to specify that all roles in the Java object jrole (resp., jrh)
will populate the EDB predicate role (resp., rh) used in the body of the first
(resp., second) ASP rule to "feed" the answer set computation. Other functions
of the challenge can be developed similarly.

The same applies to getRolesShortestPlan, the hardest function in the challenge. In particular, we drafted an encoding that combines: (*i*) *functional terms*
to encode in a uniform way different kind of updating actions given as input; (*ii*)
*arithmetic operators* to design a "temporal" encoding involving at most $2^n$ updating steps, where *n* is the number of possible actions; (*iii*) *disjunction* to guess,
at each step, the next action; (*iv*) *aggregates*, *negation* and *strong constraints*

---

[1] See http://lpop.cs.stonybrook.edu/preparing-your-position-paper

to guarantee consistency after each update; and (*v*) *weak constraints* to minimize the number of steps. A more complete description of all solutions can be downloaded from `www.mat.unical.it/ricca/downloads/LPOP-RBAC-18.zip`. Finally, database-oriented features as well as data persistence can be also added by integrating a standard RDBMS in the application. Indeed, update functions can be implemented via DML statement in SQL, and the evaluation of the prescribed constraints can be specified as ASP rules inside Java function and the execution (if needed) can be delegated to the DBMS integrating $DLV^{DB}$ [11].

**Concluding remarks.** JASP allows to combine in the same environment mainstream technologies for developing applications and ASP. This allowed us to model RBAC Challenge problems and obtain quite rapidly a prototype system.[2] Of course, obtaining a "real", complete and also efficient solution, which takes into account also other real-world nonfunctional and functional requirements, would very likely require to adopt more advanced coding strategies and additional tools (e.g., to develop a graphical interface or a WEB service).

---

[2] Two 1st year PhD students dedicated about two weeks to the project, including the time to study the challenge and the JASP framework. On the technical side they were advised by maintainer of JDLV.

**Appendix E**

# The RBAC challenge in the KB-paradigm

Marc Denecker        Jo Devriendt
Department Computer Science, KU Leuven
firstname.lastname@cs.kuleuven.be

June 25, 2018

## 1 Introduction

The RBAC challenge paper of the LPOP workshop [6] describes a dynamic system for role based access control. In this dynamic system, new users, roles and permissions are added dynamically, or existing ones are deleted. Users are assigned new roles or are stripped of them; roles are assigned new permissions or stripped of them. Roles are organized in hierarchies that may change over time. Users can pose queries. Optimized configurations of role assignments may have to be computed. Plans must be searched to realize goal configurations; selected plans must be executed. The goal of the challenge is to build a software system that implements this dynamic system and various functionalities in it.

In this position paper, we take a theoretical perspective on the problem. The questions we asked ourselves initially were of the following kind: how much of the RBAC domain can be *formally specified* in the logic FO(.) (First Order logic extended with inductive/recursive definitions and aggregates) [3]? How much of the RBAC system can be *analyzed* on the basis of the formal specification? How much of the functionalities of the RBAC system can be *executed* on the basis of the formal specification? What *forms of inference are needed* for that? Some of these questions pertain to the fundamental goals of the scientific domain of Declarative Knowledge Representation. Simple as the RBAC challenge is, we didnot know the answers at the start of this project and some questions remained unanswered at the end.

Our exercise fits in the context of what we called the *Knowledge Base Paradigm* [4, 9]. It is the idea that all problem solving is based on domain knowledge, but domain knowledge itself is inherently independent of the computational problem; formal specifications of it can be reused to solve a range of problems in the application domain. The goal of this experiment is to test this idea in the RBAC challenge: to build one formal specification, one knowledge base (or, well, as few as possible), and to reuse them in various inference problems to provide a maximal range of functionalities. We investigated how much can be *implemented/prototyped with the knowledge base system IDP* [2].

Our exercise is theoretical in the sense that we ignore the main metric in computational logic and Knowledge Representation research: efficiency/scalability. Nevertheless, we found the exercise interesting and thought provoking and we hope others will think the same. As such, we hope this paper provides some material for discussion for the LPOP workshop. New questions that we hope can be addressed during the workshop are: *where is the expertise to derive software systems from formal specifications?*, *what are the best formal specification languages for domains such as RBAC and others?*, *what are the leading systems and technologies to achieve these goals?*, *what further research is needed?*

## 2   KR: building a formal specification of RBAC

**The vocabulary**  An essential step in KR is the design of the formal vocabulary and its informal interpretation. It should be designed to express the relevant concepts of the application domain, at the right level of abstraction. In this experiment, this step was trivial since all important concepts are explicitly stated in the RBAC challenge paper [6]. The vocabulary is available in the appendix C as *vocabulary V_RBAC*.

**The theory**  We expressed the RBAC dynamic domain in the logic FO(Types,ID,Agg), as much as conveniently possible. This is First Order logic, extended with inductive/recursive definitions and agregates [7]. Below, we denote this logic as FO(.). Since FO(.) is not a temporal logic, the dynamics of the domain needs to be explicated in the vocabulary and the theory. For this we use the methodology of Linear Time Calculus (LTC) [1], a simplified version of the event calculus which uses the natural numbers as a diskrete time line. The LTC methodology introduces some fixed overhead: explicit time arguments for action and fluent (=state) symbols; frame axioms for all fluent symbols $f(ArgTypes, Time)$, expressed in terms of 3 auxiliary predicates per fluent:

- $INIT\_f(ArgTypes)$ to express the initial state of $f$;
- $C\_f(ArgTypes, Time)$ to express when $f$ is caused to be true.
- $CN\_f(ArgTypes, Time)$ to express when $f$ is caused to be false. The prefix "CN" stands for "Causes Not".

Also, standard actions of adding and deleting fluent atoms need to be specified. More than 80% of the specification is *boiler plate* overhead which in a special purpose dynamic specification language could and should be avoided.

The main components of the theory are:

- recursive definitions of fluents expressing inertia and how actions influence fluents. It contains inertia rules such as:

    $$\forall x, t : USERS(x, Next(t)) \leftarrow USERS(x, t) \land \neg CN\_USERS(x, t).$$

and causal rules such as:

$$\forall x, t : CN\_USERS(x,t) \leftarrow Delete\_USERS(x,t).$$

which expresses that the delete action causes $\neg USERS(x)$ to become true;

- action preconditions; e.g., to express that new user-role relations may be added only for active users and roles:

$$\forall x, r, t : Add\_UR(u,r,t) \Rightarrow \neg UR(u,r,t) \land USERS(u,t) \land ROLES(r,t).$$

- definitions of several derived concepts: e.g., $UserPermission$; $HUR$ which relates users with all the roles in the role hierarchy that they possess;

- concurrency axioms constraining simultaneous execution of actions.

A full theory is specified in Appendix C.

**Transactions that were not specified** There are also transactions of the RBAC software system that are not and could not (conveniently) be formalized *in* the theory. All transactions that take a *logic expression* as input were not formally specified: the operations of querying, planning, plan execution and optimisation take expressions as input. E.g., the query operation takes as input a query *expression* and returns as output the *value* of this expression in the current state. E.g., this expression could be a set-expression, and its value a set. The problem is that FO(.) lacks the expressivity to (*conveniently*) express a function from expressions to their value in the underlying structure. To express this in the logic, meta-facilities are required and they are not available in FO(.). The same argument holds for planning (the goal) and for optimisation (the cost function).

The absence of specification of these transactions in the formal theory of RBAC is a striking gap in the theory. However, it does not mean that these transactions cannot be *executed* using logical inference methods. This is discussed in the next section. But what it entails is, for instance, that any formal analysis of the RBAC theory, e.g., for proving invariants, is partial: it does not take into account the missing transactions.

**The essence of the specification** The declarative information in the RBAC challenge is quite limited. The essential information is in the definitions and in the invariants. The definitions are of the transitive closure $TransHR$ of the role hierarchy, the user roles $HUR$ in the hierarchy, and the user permissions $UserPermission$. This amounts to:

```
/* Definitions */
    { ! x[Role], y[Role], t[Time] : TransRH(x,y,t) <- RH(x,y,t).
      ! x, y, t : TransRH(x,y,t) <- ?z: RH(x,z,t) & TransRH(z,y,t).
    }
```

```
    {  ! u[User], r[Role], t[Time]: HUR(u,r,t) <-  UR(u,r,t).
       ! u, r, t: HUR(u,r,t) <- ?r1: UR(u,r1,t) & TransRH(r,r1,t).
    }

    { !u[User], p[Perm], t[Time]: UserPermission(u,p,t) <-
                                     ?r: HUR(u,r,t)) & PR(p,r,t).
    }

/* Invariants */
    ! u, r, t: UR(u,r,t) => USERS(u,t) & ROLES(r,t).

    ! p, r, t: PR(p,r,t) => PERMS(p,t) & ROLES(r,t).

    ! r, r1, t: RH(r,r1,t) => ROLES(r,t) & ROLES(r1,t).
    ! t: ~? r: TransRH(r,r,t).

    ! ssd, r, t: SSD_ROLES(ssd,r,t) => ROLES(r,t).

    ! u, ssd[SSD], t:
        #{r[Role] : HUR(u,r,t) & SSD_ROLES(ssd,r,t)} =< SSD_Card(ssd,t).
```

This theory (±) is presented in Appendix A. All the rest of the theory is boiler-place and can be generated automatically. One obtains a theory as in appendix C. Or, in a suitable dynamic logic derived from FO(.), the extra rules and assertions would be implicit in the semantics of the language. In particular, for each fluent, the frame axioms, the add and delete actions and their preconditions are similar in all cases and, in a dynamic logic, should be implicit. All functionalities specified in the RBAC challenge paper can be derived by generic inference on the completed theory. We discuss how in the next section.

**Remark** Regarding the suitability of our logic for real world applications, if we forget about the boiler plate which evidently must be eliminated, our specification is simple to understand, compact, and contains no redundancy: every aspect that was formalized needs to be formalized.

A feature of a formal specification in LTC is that it is state-oriented: transactions are atomic actions. The same is true in many dynamic specification languages. One problem that we see with, e.g., the optimization transaction or the planning transactions in RBAC is that, in practice, these operations are not atomic but they are processes involving user interaction. At the very least, the user needs to make a selection out of the possible reconfigurations or plans. Thus, a specification language may need to support the concept of process. Standard imperative or object oriented languages are strongly process oriented. Surely, this has sometimes its disadvantages as well. E.g., such process descriptions often impose unnecessary constraints on the order of execution of actions. So, one question that rises here is the issue of how to formally specify

processes, whether or when state-oriented versus process-oriented is best, and how to combine the best of worlds.

## 3  Tasks

**Analysis: Verification of invariants**   The theory `T_RBAC_Pre` in Appendix C specifies action preconditions. E.g., that a new edge $(r_1, r_2)$ can be added to the role hierarchy $RH$ only if $r_2$ is not higher in the hierarchy than $r_1$. This is to avoid that cycles are created in the hierarchy.

The RBAC challenge paper [6] mentions a set of invariants of RBAC. E.g., roles assigned to users need to be in the set $ROLES$ of active roles, and can be assigned only to elements of the set $USERS$ of active users. Others are implicit, e.g., there are no cycles in the role hierarchy. They are described above. They are the elements of the theory `invariant` in the appendix C.

One analysis task for the specification is to check if `T_RBAC_Pre` entails `invariant`. This is a deduction problem over the inductively defined set of natural numbers ($Time$). Using a standard technique, this problem can (often) be reduced to determining unsatisfiability of the following theory:

$$\texttt{T\_RBAC\_Pre\_bs} + \texttt{invariant}(0) + \neg\texttt{invariant}(1)$$

Here, `T_RBAC_Pre_bs` is the *bistate theory*, the part of theory `T_RBAC_Pre` expressing the relationship between two successive states, named $0, 1$; $invariant(0)$ and $\neg invariant(1)$ express that the invariants are satisfied at 0 and not at 1. If this theory is unsatisfiable, then any process starting from an initial state satisfying $invariant(0)$ preserves the invariants.

The theory `T_RBAC_Pre` contains an inductive definition of `TransRH` that cannot be expressed in predicate logic. To the best of our knowledge, there are currently no theorem provers for predicate logic augmented with inductive definitions[1]. The IDP system supports a light weight version: it can verify the satisfiability in the context of fixed finite domain. It is nevertheless useful. That is, the *small scope hypothesis* works fine in *many* cases: errors in the specification often do emerge in small domains.

We performed this analysis with IDP. At first, we assumed a *no concurrency* axiom, excluding the presence of multiple simultaneous actions. The analysis brought a few forgotten preconditions to the surface: namely that no element of $USERS, ROLES, PERMS$ may be be deleted when still in use in $UR, PR, RH, SSD\_Roles$.

In a second step, we analyzed concurrent execution of actions (dropping *no concurrency*). The analysis showed that with concurrency, all action preconditions need to be strengthened. So that, e.g., it is not possible to simultaneously add a role to user u and delete u of $USERS$. The action preconditions become quite complex then. However, (1) the action preconditions can be computed automatically from the invariants by the principle of **regression** [8]; (2) by adding

---

[1]Entailment of predicate logic with inductive definitions is not decidable, not even semi-decidable.

the invariants to the theory, combinations of actions that violate invariants can be detected by satisfiability checking. Thus, if the goal of an action precondition is merely to safeguard the invariants, there is no need for it: a suitable transaction engine will be able to accept or reject a proposed transaction on the basis of an LTC theory including the invariants. Thus, we can greatly simplify the theory. The appendix A contains the theory from which all boiler-plate was removed. It is the input of the IDP-solution.

We observe that not every action precondition serves to protect an invariant. E.g., an action precondition for the operation of adding $x$ to a fluent is that $x$ is not an element already of the fluent. This action precondition is not related to an invariant.

**Executing updates for RBAC** The RBAC challenge specifies a dynamic transactional system with persistent data and updates through add and delete operations. We here describe how, in theory, the updates could be derived from the formal specification.

For simplicity, we assume that times and dates are associated with natural numbers. E.g., 3:35pm on 18/7/2018 is associated with the total number of seconds that has passed since 0:00am of 1/1/1980.

Given an LTC theory $T$, we define a *state theory at time* $n \in \mathbb{N}$ as a theory consisting of the following components:

- the theory $T$, extended with

- equation $now = n$, where $now$ is a logical constant informally interpreted as the current time;

- an exhaustive description of the initial values of all fluents;

- an exhaustive description of the set of actions that occur at time points $t \leq n$ (i.e., in the past of $now$). E.g., in FO(.), the actions could be described by:

$$\left\{ \begin{array}{l} Add\_USERS(u,t) \leftarrow Future\_Add\_Users(u,t) \land t > now. \\ Add\_USERS(Jim,''2/1/2018, 10:31am''). \\ Add\_USERS(Sarah,''2/1/2018, 15:02pm''). \\ \ldots \langle \text{ set of add operations to } USERS \text{ in the past of } now \rangle \end{array} \right\}$$

This definition expresses a *local closed world assumption* on $Add\_USERS$, for the past of $now$. Here, the predicate $Future\_Add\_Users$ represents the unknown future $Add\_USERS$ transactions.

An evolution of the RBAC software system corresponds to an evolution of state theories. At each time point $n$, the state of the software system corresponds to a state theory at $n$. This state theory represents the *epistemic state* of the application: what it knows and does not know. With an update at time $n$, a new state theory corresponds which is obtained by extending the previous one with actions at time $n$. E.g., if Dave is added as a user on 18/6/2018 12am,

the above definition is extended with $Add\_USERS(Dave,"18/6/2018, 12am")$..
With time, the state theory accumulates more information about the world, in particular about the past of *now*. At no point in time, the state theory knows the future of *now*. There is a monotonicity property: the class of models/possible worlds of state theories decreases monotonically with time until, at infinity, it becomes *categorical* and has only one model: the history as it happened. Since the class of possible worlds decreases, the knowledge increases. There is one aspect that is non-monotonic though: with time, the value of *now* changes and this is a non-monotonic change. I.e., with time the application changes its mind about what is the current time. Even when nothing happens for a while, the application accumulates extra knowledge: that no change happened. Of course, the meta-operations (e.g., the queries and planning operations) are not and cannot be registered in the state theory.

Conceptually, to verify if the action preconditions and concurrency axioms are satisfied by a proposed update, the update is inserted in the state theory and the theory is verified for satisfiability. If the state theory is satisfiable, the update is accepted and the state theory is stored. Otherwise, the update is rejected and the state theory remains unchanged (except for the new value of *now*).

For a practical implementation of the above theoretical procedure, many optimizations are possible and necessary. For example, the satisfiability of the action preconditions of an update at time $now = n$ can be computed using the *current state structure*: the state at time *now*. Implementation-wise, it makes sense to explicitly store this structure. If the transaction is accepted, the current state structure can be *progressed* to the new current state [5, 1]. The current state structure is useful as well to answer what will probably be the bulk of the queries, namely queries about the current state.

**Solving current state queries and temporal queries** A state theory at time $n$ determines two structures: the *current state structure* $I_{Cur}$, expressing all fluents and actions at time $now = n$, and the *past state structure* $I_{Past}$, expressing all fluents and actions for the interval $[0, n]$. The current state structure is a structure of the *single state vocabulary*: this is the vocabulary from which time is projected out from fluents and actions.

Queries over the current state can be expressed as a set expression, or a formula or a function term in the single state vocabulary. An example is:

$$\{u[User] : \#\{p[Perm] : UserPermission(u, p)\} \geq 4\}$$

It expresses the set of users that have at least 4 permissions in the current state.

Temporal queries generalize current state queries. E.g., this current state query can be expressed as the temporal query:

$$\{u[User] : \#\{p[Perm] : UserPermission(u, p, now)\} \geq 4\}$$

Temporal queries over the past can be expressed as expressions of the same sort over the original vocabulary. E.g., the following query is whether there is a user

7
33

that once had a permission to "write", lost it and then regained it:

$$\exists u : \exists t1, t2, t3 : t1 < t2 < t3 < now \land UserPermission(u, Write, t1) \land$$
$$\neg UserPermission(u, Write, t2) \land UserPermission(u, Write, t3)$$

As explained above, the query operations cannot be formally specified in the description of the dynamic system. They can be expressed on the (procedural) meta-level and such queries can be solved by IDP in the suitable structure. Querying does not change the state and hence, it trivially preserves all invariants.

**Planning and plan execution** For this problem of the RBAC challenge, the goal is to compute a series of updates to transform the current state into a goal state satisfying a formula $\Psi[t]$. In a next phase, if the user accepts the computed plan, the plan has to be executed.

The planning inference problem takes as input the current state theory and the goal formula $\exists t : t \geq now \land \Psi[t]$. Its output is representable as an exhaustive enumeration of add and delete actions in some interval $[now, t\_end]$ so that the state theory extended with it is satisfiable and entails $\Psi[t\_end]$.

In practice, this problem can be solved using iterated model expansion. This is a well known approach in SAT for planning and in answer set programming. The search is for a model of the current state theory augmented $\Psi[now + N]$. $N$ is incremented until a model is found.

At this theoretical level, "execution" of the plan boils down to add the actions in the time interval $[now, now + N]$ in this model to the state theory. In reality, there is much more to do. E.g., execution of plans with actions that change the external world have to be monitored since they may fail. Here we will ignore this problem.

As explained above, formally specifying the planning transaction in the description of the dynamic system requires meta-facilities in the underlying logic. This does not prevent us from specifying the transaction at the procedural meta-level, using o.a. a call to a planning inference engine.

**Optimizing the configuration** The last problem of the RBAC challenge considered here is to reconfigure the base relations $UR, PR, RH$: determine minimal values for these relations such that all users maintain exactly the same permissions as in the current state.

This problem can be specified as the following inference problem. It takes as input the definition of $UserPermission/2$, the current state structure projected on the symbols $USERS, ROLES, PERMS$ and $UserPermission/2$, and finally a cost function specifying that the sum of the cardinality of the relations $UR, PR, RH$ is minimal. The output is a value for $UR, P, RH$ in a model that minimizes the cost function. This is an application of *optimization inference*.

We observe that the input of the problem contains both a value and an (inductive) definition for $UserPermission$. Thus, the model generator needs to

find values for the parameters of this definition such that the value determined by the definition corresponds to the given value.

The final step is to bring the database in the optimized state. This problem can be reduced to the application of the planning and plan execution procedure.

## 4 Implementation in IDP

We implemented a prototype system in IDP. It supports base versions of all the requested functionalities of the challenge. The system has a persistent state represented as a state theory. Implemented operations are: verification of invariants, temporal queries, updates of base relations, planning, execution of chosen plans, optimization, and planning and executing a choosen optimization. Our explicit goal was to narrowly implement the theoretical ideas, that is, to characterize and implement a maximum of functionality and flexibility on the basis of a minimal, purely declarative specification. The input of the system is the formal specification provided in the appendix A. All "boiler plate" is automatically generated from it. A trace file is presented in appendix B. None of the optimizations proposed above were implemented. The system handles only toy examples. Nevertheless, we expect that with a limited effort, it should be possible to build a system from off-the-shelve tools that can handle small applications. The system is available at `bitbucket.org/krr/rbac`.

## 5 Conclusion

The contributions of this paper are more in the scientific questions that we pose than in the complexity of the solutions that were offered. The goal of this experiment was the following: to check to what extend the RBAC software system could be implemented by generic inference on a knowledge base/formal specification. To this aim, we have evaluated the instances for RBAC of some fundamental questions of KR: what parts of the dynamic system can be formally specified, what forms of inference are needed to implement the functionalities of RBAC.

We have seen gaps in the expressivity of the logic (which occur in many, if not all current dynamic logics), namely to express complex transactions that take arguments of type *Expression* as input. That does not mean that for executing them, logic based systems are of no use. The contrary is true. However, it certainly means that the full RBAC system cannot be formally analyzed, e.g., proving invariants. It also excludes that the RBAC system as a whole can be run by uniformally applying a fixed form of inference on the specification.

All parts of the RBAC challenge can be "implemented" by inference on the formal specification(s). In all of this, the same very limited set of propositions are used time and again: the definitions of the main concepts ($UserPermission$, $HUR$, $TransRH$), concurrency axioms, invariants. Beside this, other declarative entities such as queries and goals and cost functions need to be expressed

depending of the problem at hand.

There are important functionalities that were not considered. E.g., verification of temporal logic properties. E.g., *revision inference* to erase erroneous facts. For example, assume that in 2011, a non-existing user was accidentally added and this is discovered in 2018.

# A   Input RBAC specification (without boiler plate)

```
vocabulary V{
    extern vocabulary LTCvoc

    TransRH(Role,Role,Time)
    HUR(User,Role,Time)
    UserPermission(User,Perm,Time)
}

theory Invariants:V{
    /* No concurrency axiom  */


    /* Invariants */
    !ssd,r,t: SSD_ROLES(ssd,r,t) => ROLES(r,t).

    ! u[User],ssd[SSD],t, c: SSD_Card(ssd,c,t) =>
    #{r[Role] : HUR(u,r,t) & SSD_ROLES(ssd,r,t)} =< c.
}

theory UserPermission:V{
    { ! x,y,t : TransRH(x,y,t) <- RH(x,y,t).
       ! x,y,t : TransRH(x,y,t) <- ?z: RH(x,z,t) & TransRH(z,y,t).
    }

    {  ! u,r,t: HUR(u,r,t) <-  UR(u,r,t).
       ! u,r,t: HUR(u,r,t) <- ?rr: UR(u,rr,t) & TransRH(r,rr,t).
    }

    /* UserPermission : definition:   */
    { !u,p,t: UserPermission(u,p,t) <- ?r: HUR(u,r,t) & PR(p,r,t).
    }

    /* Invariants */
    ! u,r,t: UR(u,r,t) => USERS(u,t) & ROLES(r,t).

    ! p,r,t: PR(p,r,t) => PERMS(p,t) & ROLES(r,t).

    ! r,rr,t: RH(r,rr,t) => ROLES(r,t) & ROLES(rr,t).
    !t: ~?r: TransRH(r,r,t).
}

structure S:V {
    Time = {0..20}
    now = procedure readNow
}

term ObjPlan:V{
    maxT
}

term ObjOptimize:V{
    #{u[User] r[Role] : UR(u,r,now)} +
      #{p[Perm] r[Role] : PR(p,r,now)} + #{r[Role] rr[Role] : RH(r,rr,now)}
}
```

```
vocabulary OptimizeProjection{
    extern vocabulary Types

    extern V::USERS/2
    extern V::PERMS/2
    extern V::ROLES/2
    extern V::now/0:1
    extern V::UserPermission/3
}

vocabulary OptimizeGoal{
    extern vocabulary Types

    extern V::UR/3
    extern V::PR/3
    extern V::RH/3

}
```

# B    A trace

```
$./reset.sh
$./query.sh "{ x : USERS(x,now)}"
{  }
$./query.sh "{ x : x=now}"
{ 1 }
$./query.sh "{ x : x=now}"
{ 2 }
$./update.sh "Add_USERS(u1). Add_USERS(u2). Add_USERS(u3).
Add_ROLES(r1). Add_ROLES(r2). Add_ROLES(r3).
Add_PERMS(read). Add_PERMS(write). Add_PERMS(modify).
Add_UR(u2,r2). Add_UR(u3,r3).
Add_PR(write,r1). Add_PR(read,r2). Add_PR(modify,r3).
Add_RH(r1,r2). Add_RH(r2,r3)."
Update succesful.
$./query.sh "{ x p : UserPermission(x,p,now)}"
{ u2,read; u2,write; u3,modify; u3,read; u3,write }
$./update.sh "Add_USERS(u1)."
Warning: given update violates preconditions and is aborted.
$./update.sh "Delete_USERS(u1)."
Update succesful.
$./update.sh "Add_USERS(u1)."
Update succesful.
$./update.sh " Add_RH(r3,r3)."
Warning: given update violates preconditions and is aborted.
$./update.sh " Add_RH(r3,r2)."
Warning: given update violates preconditions and is aborted.
$./query.sh "{ x p : TransRH(x,p,now)}"
{ r1,r2; r1,r3; r2,r3 }
$./plan.sh "/*all users have all permissions; only one action per time point */
! x[User], p1[Perm]: UserPermission(x,p1,maxT)  &
 ! t: t>= now => #{u:Add_USERS(u,t)}+
    #{u:Add_ROLES(u,t)}+
    #{u:Add_PERMS(u,t)}+
    #{u, r:Add_UR(u,r,t)}+
```

```
        #{p, r :Add_PR(p,r,t)}+
        #{r,rr :Add_RH(r,rr,t)}+
        #{s,r :Add_SSD_ROLES(s,r,t)}+
        #{s,c :Add_SSD_Card(s,c,t)}+
        #{u:Delete_USERS(u,t)}+
        #{u:Delete_ROLES(u,t)}+
        #{u:Delete_PERMS(u,t)}+
        #{u, r:Delete_UR(u,r,t)}+
        #{p, r :Delete_PR(p,r,t)}+
        #{r,rr :Delete_RH(r,rr,t)}+
        #{s,r :Delete_SSD_ROLES(s,r,t)}+
        #{s,c :Delete_SSD_Card(s,c,t)}=<1. "
A correct plan is:
1) Add_PR(modify,r2).
2) Add_UR(u1,r3).
Commit this plan? (y/n/q) y
$./optimize.sh
Optimization:
PR(modify,r2). PR(read,r2). PR(write,r2).
UR(u1,r2). UR(u2,r2). UR(u3,r2).

Proposed plan:
1) Add_PR(write,r2). Delete_PR(modify,r3).  Delete_PR(write,r1).
Delete_RH(r1,r2). Delete_RH(r2,r3). Add_UR(u3,r2).
Delete_UR(u3,r3).
Commit this plan? (y/n/q) y
$
```

## C    RBAC with action preconditions and without concurrency

```
LTCvocabulary V_RBAC{
    type Time isa int
    Start:Time
    partial Next(Time):Time

    type User
    type Role
    type Perm

    USERS(User,Time)
    Add_USERS(User,Time)
    Delete_USERS(User,Time)

    Init_USERS(User)
    C_USERS(User,Time)
    CN_USERS(User,Time)

    UsedUSERS(User,Time)
    UsedROLES(Role,Time)
    UsedPERMS(Perm,Time)

    ROLES(Role,Time)
    Add_ROLES(Role,Time)
    Delete_ROLES(Role,Time)
```

```
Init_ROLES(Role)
C_ROLES(Role,Time)
CN_ROLES(Role,Time)

PERMS(Perm,Time)
Add_PERMS(Perm,Time)
Delete_PERMS(Perm,Time)

Init_PERMS(Perm)
C_PERMS(Perm,Time)
CN_PERMS(Perm,Time)

UR(User,Role,Time)
Add_UR(User,Role,Time)
Delete_UR(User,Role,Time)

Init_UR(User,Role)
C_UR(User,Role,Time)
CN_UR(User,Role,Time)

PR(Perm,Role,Time)
Add_PR(Perm,Role,Time)
Delete_PR(Perm,Role,Time)

Init_PR(Perm,Role)
C_PR(Perm,Role,Time)
CN_PR(Perm,Role,Time)

UserPermission(User,Perm,Time)

/* hierarchical */
RH(Role,Role,Time)
Add_RH(Role,Role,Time)
Delete_RH(Role,Role,Time)

Init_RH(Role,Role)
C_RH(Role,Role,Time)
CN_RH(Role,Role,Time)

TransRH(Role,Role,Time)
HUR(User,Role,Time)   /* AuthorizedRole */
HUserPermission(User,Perm,Time)

/*SSD*/

type SSD
type NrRoles isa int
SSD_ROLES(SSD,Role,Time)
Add_SSD_ROLES(SSD,Role,Time)
Delete_SSD_ROLES(SSD,Role,Time)

Init_SSD_ROLES(SSD,Role)
C_SSD_ROLES(SSD,Role,Time)
CN_SSD_ROLES(SSD,Role,Time)

SSD_Card(SSD,Time):NrRoles
Set_SSD_Card(SSD,NrRoles,Time)
```

```
         Init_SSD_Card(SSD,NrRoles)
         C_SSD_Card(SSD,NrRoles,Time)

}

theory T_RBAC_Pre: V_RBAC{

    { ! u,t:UsedUSERS(u,t)<- ?r:UR(u,r,t).}

    { ! r,t:UsedROLES(r,t)<- ?u:UR(u,r,t).
      ! r,t:UsedROLES(r,t)<- ?p:PR(p,r,t).
       ! r,t:UsedROLES(r,t)<- ?r1:RH(r,r1,t).
       ! r,t:UsedROLES(r,t)<- ?r1:RH(r1,r,t).
       ! r,t:UsedROLES(r,t)<-? ssd: SSD_ROLES(ssd,r,t). }

    { ! p, t :UsedPERMS(p,t)<- ?r:PR(p,r,t).}

 /*RBAC pure */
    { !u: USERS(u,Start) <- Init_USERS(u).
      !u, t: USERS(u,Next(t)) <- C_USERS(u,t).
      !u, t: USERS(u,Next(t)) <- USERS(u,t) & ~CN_USERS(u,t).

      !u,t: C_USERS(u,t) <- Add_USERS(u,t).
      !u,t: CN_USERS(u,t) <- Delete_USERS(u,t).
    }

    ! x,t: Add_USERS(x,t) => ~USERS(x,t).
    ! x,t: Delete_USERS(x,t) => USERS(x,t) & ~ UsedUSERS(x,t).

    { !r: ROLES(r,Start) <- Init_ROLES(r).
      !r, t: ROLES(r,Next(t)) <- C_ROLES(r,t).
      !r, t: ROLES(r,Next(t)) <- ROLES(r,t) & ~CN_ROLES(r,t).

      !r,t: C_ROLES(r,t) <- Add_ROLES(r,t).
      !r,t: CN_ROLES(r,t) <- Delete_ROLES(r,t).
    }

    ! x,t: Add_ROLES(x,t) => ~ROLES(x,t).
    ! x,t: Delete_ROLES(x,t) => ROLES(x,t) & ~ UsedROLES(x,t).


    { !p: PERMS(p,Start) <- Init_PERMS(p).
      !p, t: PERMS(p,Next(t)) <- C_PERMS(p,t).
      !p,t: PERMS(p,Next(t)) <- PERMS(p,t) & ~CN_PERMS(p,t).

      !p,t: C_PERMS(p,t) <- Add_PERMS(p,t).
      !p,t: CN_PERMS(p,t) <- Delete_PERMS(p,t).
    }

    ! x,t: Add_PERMS(x,t) => ~PERMS(x,t).
    ! x,t: Delete_PERMS(x,t) => PERMS(x,t)& ~ UsedPERMS(x,t).

    { !u,r: UR(u,r,Start) <- Init_UR(u,r).
      !u, r, t: UR(u,r,Next(t)) <- C_UR(u,r,t).
      !u,r,t: UR(u,r,Next(t)) <- UR(u,r,t) & ~CN_UR(u,r,t).
```

```
        !u,r,t: C_UR(u,r,t) <- Add_UR(u,r,t).
        !u,r,t: CN_UR(u,r,t) <- Delete_UR(u,r,t).
      }

      ! u,r,t: Add_UR(u,r,t) => ~UR(u,r,t).
      ! u,r,t: Add_UR(u,r,t) => USERS(u,t) & ROLES(r,t).
      ! u,r,t: Delete_UR(u,r,t) => UR(u,r,t).

      { !p,r: PR(p,r,Start) <- Init_PR(p,r).
        !p, r, t: PR(p,r,Next(t)) <- C_PR(p,r,t).
        !p,r,t: PR(p,r,Next(t)) <- PR(p,r,t) & ~CN_PR(p,r,t).

        !p,r,t: C_PR(p,r,t)<-Add_PR(p,r,t).
        !p,r,t: CN_PR(p,r,t)<-Delete_PR(p,r,t).
      }

      ! p,r,t: Add_PR(p,r,t) => ~PR(p,r,t).
      ! p,r,t: Add_PR(p,r,t) => PERMS(p,t) & ROLES(r,t).
      ! p,r,t: Delete_PR(p,r,t) => PR(p,r,t).

      { !u,p,t: UserPermission(u,p,t) <- ?r: UR(u,r,t) & PR(p,r,t).
      }

/* Hierarchical */

      { !r,r1: RH(r,r1,Start) <- Init_RH(r,r1).
        !r,r1, t: RH(r,r1,Next(t)) <- C_RH(r,r1,t).
        !r,r1, t: RH(r,r1,Next(t)) <- RH(r,r1,t) &  ~CN_RH(r,r1,t).

        !r,r1,t: C_RH(r,r1,t)<-Add_RH(r,r1,t).
        !r,r1,t: CN_RH(r,r1,t)<-Delete_RH(r,r1,t).
      }

      ! r,r1,t: Add_RH(r,r1,t) => ~RH(r,r1,t).
      ! r,r1,t: Add_RH(r,r1,t) => ROLES(r,t) & ROLES(r1,t) & r~=r1.
      ! r,r1,t: Add_RH(r,r1,t) => ~TransRH(r1,r,t).
      ! r,r1,t: Delete_RH(r,r1,t) => RH(r,r1,t).

      { ! x,t : TransRH(x,y,t) <- RH(x,y,t).
        ! x,y,t : TransRH(x,y,t) <- ?z: RH(x,z,t) & TransRH(z,y,t).
      }

      { ! u,r,t: HUR(u,r,t) <- ?r1: UR(u,r1,t) & TransRH(r,r1,t).
      }

      { !u,p,t: HUserPermission(u,p,t)<- ?r: (UR(u,r,t) | HUR(u,r,t)) & PR(p,r,t).
      }

/* SSD */
      { !ssd,r: SSD_ROLES(ssd,r,Start) <- Init_SSD_ROLES(ssd,r).
        !ssd,r, t: SSD_ROLES(ssd,r,Next(t)) <- C_SSD_ROLES(ssd,r,t).
        !ssd,r,t: SSD_ROLES(ssd,r,Next(t)) <- SSD_ROLES(ssd,r,t) & ~CN_SSD_ROLES(ssd,r,t).

        !ssd,r,t: C_SSD_ROLES(ssd,r,t) <- Add_SSD_ROLES(ssd,r,t).
        !ssd,r,t: CN_SSD_ROLES(ssd,r,t) <- Delete_SSD_ROLES(ssd,r,t).
      }
```

```
    ! ssd,r,t: Add_SSD_ROLES(ssd,r,t) => ~SSD_ROLES(ssd,r,t).
    ! ssd,r,t: Delete_SSD_ROLES(ssd,r,t) => SSD_ROLES(ssd,r,t).

    { !ssd,c: SSD_Card(ssd,Start)=c <- Init_SSD_Card(ssd,c).
      !ssd,c, t: SSD_Card(ssd,Next(t))=c <- C_SSD_Card(ssd,c,t).
      !t,ssd: SSD_Card(ssd,Next(t))=SSD_Card(ssd,t) <- ~? c: C_SSD_Card(ssd,c,t).

      !t,ssd,c: C_SSD_Card(ssd,c,t)<- Set_SSD_Card(ssd,c,t).
    }

    ! u[User],ssd[SSD],t: #{r[Role] : HUR(u,r,t) & SSD_ROLES(ssd,r,t)} =< SSD_Card(ssd,t).

 //No concurrency
   ! t: #{u:Add_USERS(u,t)}+
#{u:Add_ROLES(u,t)}+
#{u:Add_PERMS(u,t)}+
#{u, r:Add_UR(u,r,t)}+
#{p, r :Add_PR(p,r,t)}+
#{r,r1 :Add_RH(r,r1,t)}+
#{s,r :Add_SSD_ROLES(s,r,t)}+
#{s,c :Set_SSD_Card(s,c,t)}+
#{u:Delete_USERS(u,t)}+
#{u:Delete_ROLES(u,t)}+
#{u:Delete_PERMS(u,t)}+
#{u, r:Delete_UR(u,r,t)}+
#{p, r :Delete_PR(p,r,t)}+
#{r,r1 :Delete_RH(r,r1,t)}+
#{s,r :Delete_SSD_ROLES(s,r,t)}=<1.

}

theory invariant: V_RBAC{
    ! u,r,t: UR(u,r,t) => USERS(u,t) & ROLES(r,t).
    ! p,r,t: PR(p,r,t) => PERMS(p,t) & ROLES(r,t).

    ! r,r1,t: RH(r,r1,t) => ROLES(r,t) & ROLES(r1,t).
    !t: ~?r: TransRH(r,r,t).

    !ssd,r,t: SSD_ROLES(ssd,r,t) => ROLES(r,t).

    ! t[Time]: !u[User],ssd[SSD]:
            #{r[Role] : HUR(u,r,t) & SSD_ROLES(ssd,r,t)}
                                        =< SSD_Card(ssd,t).
}
```

# Appendix F

# The RBAC challenge in LogiQL: Solutions and Limitations

K. Tuncay Tekle, Stony Brook University
tuncay@cs.stonybrook.edu

June 28, 2018

## 1 Introduction

LogicBlox [1] is a commercial system unifying the programming model for enterprise software development that combines transactions with analytics by using a flavor of Datalog called LogiQL. LogiQL is a strongly-typed, extended form of Datalog that allows coding of entire enterprise applications, including business logic, workflows, user interfaces, statistical modeling, and optimization tasks.

The programming challenge for the Logic and Practice of Programming workshop is based on Role-Based Access Control (RBAC) [2], and provides various problems whose solutions require the handling of (1) rules, (2) queries, (3) updates, (4) type constraints, (5) other constraints, (6) optimization, (7) planning. In this paper, we show that problems related to the first five can be completely handled, one case of optimization can be handled with some caveats, and another can be written but does not run in the latest implementation; and planning cannot be handled in LogiQL.

As an overview, LogicBlox is a state-based system with a persistent database that can be manipulated, where one can add facts to the database, and rules and constraints to the state, and query the database at any point in time. Rules and queries are readily handled by a logic programming system as expected. Type constraints are also intrinsic to LogiQL as every predicate is typed. Constraints that are not type constraints in this challenge can also be intuitively expressed. All of these are maintained in the presence of updates. Optimization problems can also be handled, although we show that it requires a rewrite to fit in the optimization framework, and has some restrictions. There is no capability for planning in LogiQL, however we propose a syntax similar to optimization that requires an implementation. However, since LogicBlox is a closed-source system, we do not expect that this would be implemented.

## 2 Implementation

We briefly introduce how LogicBlox works, and how it evaluates LogiQL. In LogicBlox, the system starts with an empty state, where a state contains the fact database, rules and constraints. The state can be manipulated by adding facts, rules or constraints, where more facts can be added automatically by the rules due to the addition of facts, or the manipulation can be rejected due to the added facts not satisfying constraints.

The programming system can be separated into three components: (1) installed blocks, (2) updates, (3) queries. Installed blocks contain entity and predicate declarations, rules, and constraints. Updates are inserts, updates or deletes to entities or predicates. Queries retrieve data from predicates or may introduce a temporary rule to retrieve data. A sketch of how LogicBlox evaluates LogiQL as follows: (1) when a new block is installed, all declarations in the block are installed, all rules in the block are evaluated bottom-up with respect to the current state of the database, and all constraints are checked; if a constraint fails, the installation of the block is rolled back; (2) for an update: all rules in all installed blocks are evaluated bottom-up (incrementally whenever possible), then all type constraints are enforced by deleting facts of a predicate related to an entity if that entity is deleted in an update, then if any other constraint is not satisfied after the evaluation of the rules, the update is rolled back; (3) for a query, the results of the query are retrieved as expected.

Next, we dig in to the components of RBAC, where we talk about only new concepts needed for each component in order. The code for each component can be found in the appendix.

## 2.1 Core RBAC

**Entity declarations and updates.** This component first defines some types: Users, roles, and permissions. Types are called entities in LogiQL, and entities can be implemented via *entity* and *refmode* (reference mode) declarations. A user can then be added using update statements. All of the update functions for these entities (`AddUser, DeleteUser, AddRole, DeleteRole, AddPerm, DeletePerm`) can be implemented using the entity declarations and update statements.

**Relation declarations and updates.** Next, user-role pairs, and permission-role pairs are introduced with the constraint that each set of pairs is a subset of the cross-product of the entities. This is trivially satisfied in LogiQL by the nature of the declaration of the set of pairs, where each argument's type is provided by a constraint. Therefore, e.g., any insert to a predicate verifies that the types of its arguments matches the declaration, or else the update fails. If an entity of a type is removed, all predicate facts related to that particular entity is removed as well. All of the update functions for these relations (`AddUR, DeleteUR, AddPR, DeletePR`) can be implemented using these declarations.

**Queries.** All three queries in this component are easily implemented via logical queries.

## 2.2 Hierarchical RBAC

**Rules.** Reflexive-transitive closure of the role hierarchy can be defined via rules.

**Complex constraints.** So far we have only seen type constraints. More complex constraints can be similarly implemented via the right arrow notation. For example, the acyclicity of the role hierarchy can be enforced by constraints.

## 2.3 Static Separation of Duty (SSD)

SSD can be implemented using the concepts introduced above, and the hieararchical version only needs to change the predicate to use the Hierarchical RBAC version rather than the Core.

## 2.4 Administrative RBAC

There are two types of new challenges in this component: optimization and planning. There is no planning functionality in LogiQL, and optimization has restrictions. We show how to solve the `MinRoleAssignments` and `MinRoleAssignmentsWithHierarchy` optimization problems with caveats, and suggest a syntax for planning.

**Optimization.** In LogiQL, optimization is performed by translating the rules and variables into a mathematical program. Therefore, the variables and the predicate to solve for need to be defined, and there are various restrictions on what the rules and constraints can look like. Then, the mathematical problem is solved via an invocation of a solver using the facts of the current state. Note that this is significantly different than how the system normally operates.

For the hierarchical version of the problem, the definition of rules needs to be recursive. However, the preprocessor fails at translating the rules written this way, although a mapping seems obvious for this example.

**Planning.** Planning problems in Administrative RBAC cannot be specified in LogiQL, but the lack of an implementation (and the possibility of it being implemented being close to zero) notwithstanding, it is not difficult to imagine that directives such as the ones used in optimization could be used to construct action variables based on each action that can be taken, and a solver variable for the objective.

# A   Core RBAC

- Definition of Users, Roles, Permissions (types):

  ```
  User(u), UserName(u:n) -> string(n).
  Role(r), RoleName(r:n) -> string(n).
  Permission(p), PermissionName(p:n) -> string(n).
  ```

- Addition/removal of a user (update):

  ```
  +UserName[_] = "tuncay".
  -UserName[_] = "tuncay".
  ```

- Definition of user-role, permission-role pairs (relations):

  ```
  UR(u,r) -> User(u), Role(r).
  PR(p,r) -> Permission(p), Role(r).
  ```

- Addition/deletion of user-role, permission-role pairs (update):

  ```
  +UR(u,r) <- UserName[u] = "tuncay", RoleName[r] = "admin".
  -PR(p,r) <- PermissionName[u] = "write", RoleName[r] = "admin".
  ```

- `AssignedRoles`, `UserPermissions`, `CheckAccess` queries:

  ```
  _(r) <- UserName[u] = "tuncay", UR(u,r).
  _(p) <- UserName[u] = "tuncay", UR(u,r), PR(p,r).
  _(true) <- UserName[u] = "tuncay", PermissionName[p] = "read", UR(u,r), PR(p,r).
  ```

# B   Hierarchical RBAC

- Role hierarchy definition:

  ```
  RH(r1,r2) -> Role(r1), Role(r2).
  ```

- Definition of the reflexive-transitive closure of the role hierarchy, and `AuthorizedRoles` (rules):

  ```
  Trans(r1,r1) <- Role(r1).
  Trans(r1,r2) <- RH(r1,r2).
  Trans(r1,r2) <- RH(r1,r3), Trans(r3,r2).

  AuthorizedRoles(u,r1) <- UR(u,r), Trans(r,r1).
  ```

- Enforcement of the acyclicity of the role hierarchy (constraint):

  ```
  Trans(p,p2), Trans(p2,p) -> p = p2.
  ```

# C  Static Separation of Duty

- Declarations of types and relations:

  ```
  RoleSet(rs), RoleSetName(rs:n) -> string(n).
  RoleSetRoles(rs,r) -> RoleSet(rs), Role(r).
  SSDItem(s), SSDItemName(s:n) ->string(n).
  SSDItemRoleSet[s] = rs -> SSDItem(s), RoleSet(rs).
  SSDItemCardinality[s] = c -> SSDItem(s), int(c).
  SSDItem(s) -> SSDItemRoleSet[s] = _.
  SSDItem(s) -> SSDItemCardinality[s] = _.
  ```

- Deletion of an SSDItem:

  ```
  -SSDItem[s], -RoleSet(rs) <- SSDItemName@prev[s] = "s1", SSDItemRoleSet@prev[s] = rs.
  ```

  Note here the use of @prev on the right hand side of the rule. For an update of a predicate that depends on a fact of that predicate before the update starts, this annotation is utilized to avoid recursion. So this rule says if there is an SSD item with the name s1 before the update is started, then delete that SSD item.

- Adding/deleting role sets, updating cardinality:

  ```
  +RoleSetRoles(rs,r) <- SSDItemName[s] = "s1", SSDItemRoleSet[s] = rs.
  -RoleSetRoles(rs,r) <- SSDItemName[s] = "s1", SSDItemRoleSet[s] = rs, RoleName[r] = "r1".
  ^SSDItemCardinality[s] = c <- SSDItemName[s] = "s1".
  ```

  The ^ symbol indicates an update of an existing functional value. We need to define the number of items in a role set to support the constraints as follows.

- Constraints for the role count and cardinality relations:

  ```
  RoleCount[s] = rc -> SSDItem(s), int(rc).
  RoleCount[s] = rc <- agg<<rc = count()>>
                       SSDItemRoleSet[s] = rs, RoleSetRoles(rs,_).

  SSDItemCardinality[s] = c -> c > 0, c < RoleCount[s].

  UserSSDRoleCount[u,s] = rc -> User(u), SSDItem(s), int(rc).
  UserSSDRoleCount[u,s] = rc <- agg<<rc = count()>>
                                UR(u,r), SSDItemRoleSet[s] = rs, RoleSetRoles(rs,r).
  UserSSDRoleCount[_,s] < SSDItemCardinality[s].
  ```

- Queries:

  ```
  _(name) <- SSDItemName[_] = name.
  _(r) <- SSDItemName[s] = "s1", SSDItemRoleSet[s] = rs, RoleSetRoles(rs,r).
  _(c) <- SSDItemName[s] = "s1", SSDItemCardinality[s] = c.
  ```

- For Hierarchical RBAC, the only necessity is to change the rule defining UserSSDRoleCount to use AuthorizedRoles instead of UR.

# D  Administrative RBAC

For the first problem, `MinRoleAssignments`, the total size of new `UR'` and `PR'` need to be minimized; however `ROLES'` can also contain new roles. There is no way to invent new values during optimization in LogiQL. Therefore, we restrict the question so that `ROLES'=ROLES`. Then, we first define two predicates which define whether a role subsumes another and the original permissions.

```
Subsume[r1,r2] = b -> Role(r1), Role(r2), int(b).
Subsume[r1,r2] = 1 <- Trans(r1,r2).
OrigPerms[u,p] = b -> User(u), Permission(p), int(b).
OrigPerms[u,p] = 1 <- AuthorizedRoles(u,r), PR(r,p).
```

Then, we define the new variables and constraints for the optimization problem as follows, where `NewPerms` are the user-permissions based on the `NewUR` and `NewPR` sets which are self-explanatory:

```
NewUR[u,r] = b -> User(u), Role(r), int(b), b >= 0, b <= 1
NewPR[p,r] = b -> Permission(p),  Role(r), int(b), b >= 0, b <= 1.
lang:solver:variable(`NewUR).
lang:solver:variable(`NewPR).

NewPerms[u,p] = b -> User(u), Permission(p), int(b), b >= 0, b <= 1.
NewPerms[u,p] = b1, OrigPerms[u,p] = b2 -> b1 = b2.
NewPerms[u,p] = b1, !OrigPerms[u,p] = _ -> b1 = 0.
NewPerms[u,p] = v, NewUR[u,r1] = v1, Subsume[r1,r2] = v2, NewPR[p,r2] = v3 -> v >= v1 * v2 * v3.
```

The directive `lang:solver:variable` defines the variables of the optimization problem. Finally, we define the optimization function as follows:

```
NewURSize[] = nus <- agg<<nus=total(v)>>
                    v = NewUR[_,_].
NewPRSize[] = nps <- agg<<nps=total(v)>>
                    v = NewPR[_,_].
TotalSize[] = tc -> int(tc).
TotalSize[] = NewURSize[] + NewPRSize[].
lang:solver:minimal(`TotalSize).
```

The directive `lang:solver:minimal` defines the predicate to solve for (by way of minimization).

# Appendix G
# Logic-based Methods for Software Engineers and Business People

Joost Vennekens

KU Leuven, Dept. of Computer Science
Campus De Nayer, St-Katelijne-Waver, Belgium

**Abstract.** Both software engineers and business people tend to be reluctant to adopt logic-based methods. On the one hand, this may be due to unfamiliarity with "scary" logical syntax. To ameliorate this, we developed an API for a state-of-the-art logic system, using only standard Python syntax. On the other hand, logic-based methods might have more impact if they could be used directly by business people. The recent DMN standard that might help in this respect.

Logic-based methods have great potential to improve current software engineering practice. However, this potential is often not evident for people from industry. For instance, when demoing a state-of-the-art knowledge base system to programmers, we typically get comments stating that, e.g., they can program the same functionality with "a couple of for-loops". To properly appreciate the benefits of a logic-based approach (greater flexibility, modularity, reliability and maintainability), it is necessary to understand in detail how the representation works, at least in the context of a small example. Here, syntax often seems a bottleneck, with programmers getting hung up on details they dislike.

To reduce this effect, we developed an API [3] that allows the state-of-the-art IDP knowledge base system [1] to be used from the Python programming language. Crucially, this API allows logic formulas to be added to the knowledge base using *standard Python syntax*. For instance, suppose that a mapping from users to roles is given by the following Python data structure, consisting of a list of tuples, and that a list of access rights is given in a similar way:

$$InRole = [\ ('John', 'Student'), ('Ann', 'Admin')\ ] \quad (1)$$
$$Allowed = [\ ('Student', 'PublicData'), ('Admin', 'Passwords')\ ] \quad (2)$$

Suppose we now want to verify that certain users indeed have access to certain resources, e.g., $Access = [\ ('Ann', 'Passwords')]$. The following Python expression then checks that Ann is indeed allowed to access the password data:

all(any((u,r) in InRole and (r,res) in Allowed for r in Role) for (u,res) in Access)

Our API accepts the same Python expression, but allows it to be used in different ways: for instance, if the programmer does not assign a value to *InRole* herself (i.e., she omits statement (1)), then the IDP system will itself compute

an assignment of users to roles that ensures that all of the required accesses are possible (i.e., *Ann* will be placed in the role *Admin* to ensure that she can access the password data). In this way, the flexibility of a declarative approach can be experienced without any new syntax. We therefore believe that this API may be a useful tool to introduce programmers to the power of logic-based approaches.

In many contexts, however, not programmers need to be convinced, but business people. Despite the significant technical differences between traditional software engineering and knowledge-based methods, business users may simply see a choice between either paying a programmer to deliver a piece of software or paying a knowledge engineer to deliver a piece of software. Here, we believe that the unique selling proposition of knowledge-based methods is that they may allow to "eliminate the middle man", by giving ownership of the domain knowledge back to the business instead of to an IT departement.

The recent Decision Model and Notation (DMN) standard[1] published by the Object Management Group has been developed specifically to allow domain experts without any background in logic or computer science to construct a formal model of a decision process. Using this notation, a business expert could define the decision logic for access control by the following three tables:

| **C** | Input | Output |
|---|---|---|
|  | User | Role |
|  | "John" | "Student" |
|  | "Ann" | "Admin" |
|  | "Ann" | "Student" |

| **C** | Input | Output |
|---|---|---|
|  | Role | Resource |
|  | "Student" | "PublicData" |
|  | "Admin" | "PublicData" |
|  | "Admin" | "Passwords" |

| **U** | Input | Output |
|---|---|---|
|  | Role | AccessGranted *default: false* |
|  | Request | true |

Here, the "C" in the top left of the first table represents the "Collect" hit policy, meaning that a user belongs to *all* of the roles that are mentioned for her in the table. The second table expresses that the user has access to all resources that correspond to one of the roles to which she belongs. Finally, the third table expresses that when a user requests a resource to which she has access, the access should be granted, and otherwise it should not.

The tabular notation is not only intuitive for business people, it also allows the completeness and correctness of the decision procedure to be easily checked. Currently, most reasoning systems for DMN have evolved from rule-based expert systems and they typically only allow forward propagation. However, since a DMN model expresses a purely declarative piece of knowledge, there is no reason why other inference tasks could not be applied to it, for instance, by translating such a model to input for the IDP system, as done (currently still manually) in [2]. We suspect that the DMN language might easily be extendable to allow more complex kinds of knowledge to be expressed, while retaining the ease-of-use for domain experts. In this way, more of the power of logic-based systems could be put directly at the finger tips of domain experts and business decision makens, eliminating the need for an intervening programmer or knowledge engineer.

---

[1] https://www.omg.org/spec/DMN/

# Appendix H

# Easier Rules and Constraints for Programming[*]

Yanhong A. Liu    Scott D. Stoller

We discuss how rules and constraints might be made easier for more conventional programming. We use a language that extends DistAlgo, which extends Python, and we use the RBAC programming challenge plus distributed RBAC as examples.

**Python.** Python is a high-level programming language with an easy-to-read syntax. It supports conventional imperative programming and object-oriented programming. It also supports database-style programming with sets and queries (comprehension and generator expressions) and functional programming with recursive functions and even a syntax for lambda. However, it does not support rules and constraints.

**DistAlgo.** DistAlgo [LSL17, LL17] is a language that extends Python to support distributed programming with processes and message passing. It also extends Python to support more powerful queries with constraints and tuple patterns, including logic quantifications with witnesses. These query constructs were created to better express high-level synchronization conditions over messages and processes but also high-level queries in general, while integrating seamlessly with imperative programming.

For example, consider a set `UR` of user-role pairs and a particular user `user`. The set of roles that `user` has can be expressed using a set comprehension with a tuple pattern as follows.

```
setof(r, (_user,r) in UR)
```

The membership condition is exactly a constraint, and in general any number of constraints can be used. `(_user,r)` is a tuple pattern, where the underscore indicates a variable on the left side of a membership clause whose value is bound before the query. Note that we have also implemented a more ideal syntax for the same query, shown below, but here we use Python accepted syntax, shown above, so that the Python parser can be used.

```
{r: (=user,r) in UR}
```

Similarly one may compute aggregation (e.g., `countof` and `minof`) over sets, and universal and existential quantifications (`each(x in s, has= p(x)` or `some(x in s, has= p(x))`).

**Extension with constraint optimization.** With the more powerful set queries as above, it is easy to write an additional constraint to filter out only those that minimize some objective function, e.g., the constraint `f(r) == minof(f(x), (_user,x) in UR)` can be inserted in the set comprehension shown above. It is even easier to simply add the constraint as follows,

```
minimize= exp
```

where `exp` expresses the objective function, e.g., `minimize= f(r)` can be inserted in the set comprehension shown above. This is just as in mathematical programming tools.

**Extension with rules.** Just as declaring a named function or method, one should be able to easily declare a named set of rules, e.g.,

```
def rules (name='Trans_rules'):
  if edge(x,y): path(x,y)
  if edge(x,z) and path(z,y): path(x,y)
```

and call an inference function to infer values using the rule set, e.g., the following returns the set of pairs for which predicate `path` holds using rule set `Trans_rules` given a set of edges `RH`.

```
infer(path, edge=RH, rules=Trans_rules)
```

One can also use `path(1,v)` in place of `path` to return the set of values of `v` for which `path(1,v)` holds. Note that predicates `edge` and `path` are simply set-valued variables, without needing high-order logic.

**Extension with backtracking under choices.** While planning problems can be expressed as constraint solving and optimization, it is more direct if actions in the program can be expressed with choices, with actions sequenced, with backtracking in an allowed scope, until sequences of actions satisfying a condition are found and returned or all choices are enumerated. This is easily expressed with a pair of `assume` and `achieve` statements that surround statements with choices.

In particular, given a set of actions `acts` that are allowed operations, e.g., method definitions, let `instances(acts)` generate all instances of calls to those methods, and let `do(a)` execute method call `a`. The following code finds any sequence that satisfies `condition`, where `some` makes a choice.

```
assume(True)
seq = []
while not condition:
  if some(a in instances(acts)):
    do(a)
    seq.append(a)
achieve(anyof(seq))
```

Also, a cost function can be computed along the sequence, and solutions that minimize the cost may be returned.

**Implementation.** The extensions are being implemented by extending DistAlgo. The implementation is currently incomplete. The main challenge will be efficient implementation to provide competitive performance compared with lower-level or more complex manually programmed solutions.

**RBAC programming challenge solution.** The Appendix shows how to express all components and functions of the RBAC programming challenge, plus a component for distributed RBAC, in the extended language. It is aimed to express everything in the clearest and most direct way possible.

`process` in class header is needed only for distributed execution for the distributed RBAC component; for others, it is included only to allow use of more powerful set queries with constraints and tuple patterns. `pre` for preconditions could be implemented simply as `assert` in Python but we plan to support it directly in the extensions to DistAlgo.

These components can run with DistAlgo: `CoreRBAC`, `HierarchicalRBAC_set`, `HierarchicalRBAC`, `CoreRBACwithSSD`, `HierarchicalRBACwithSSD`, and `DistRBAC`. Their less powerful variants in Python without `DistRBAC` were run and optimized to run efficiently previously [LWG+06, GLSR12].

These do not currently run: `HierarchicalRBAC_rules` and `AdminRBAC`.

# Appendix: RBAC challenge in a language that extends Python

```python
"""
We consider Role-Based Access Control (RBAC) with 6 components:

  Core RBAC,
  Hierarchical RBAC,
  Core RBAC with Static Separation of Duty constraint (also called Constrained RBAC),
  Hierarchical RBAC with Static Separation of Duty constraint,
  Administrative RBAC, and
  Distributed RBAC
"""

class CoreRBAC(process):
  """
  Core RBAC keeps several sets including the following:

    USERS: set of users
    ROLES: set of roles
    PERMS: set of permissions
    UR: set of user-role pairs
    PR: set of permission-role pairs

  with constraints:

    UR subset USERS * ROLES
    PR subset PERMS * ROLES

  update functions for each set, subject to the constraints above:

    AddUser, DeleteUser, AddRole, DeleteRole, AddPerm, DeletePerm
    AddUR, DeleteUR, AddPR, DeletePR
    each Add has pre-conditions:
      the element is not yet in the set and the constraints will not be violated
    each Delete has the pre-condition that the element is in the set,
      and maintains the constraints
```

```
36      query functions including the following:
37
38         AssignedUsers(role): the set of users assigned to role in UR
39         AssignedRoles(user): the set of roles assigned to user in UR
40         UserPermissions(user):
41            the set of permissions assigned to the roles assigned to user
42         CheckAccess(user, perm):
43            whether some role is assigned to user and is granted perm
44      """
45
46      def setup():
47         self.USERS = set()
48         self.ROLES = set()
49         self.PERMS = set()
50         self.UR = set()              # UR subset USERS * ROLES
51         self.PR = set()              # PR subset PERMS * ROLES
52
53      def AddUser(user):              # pre: user not in USERS
54         USERS.add(user)
55
56      def DeleteUser(user):           # pre: user in USERS
57         UR -= setof((user,r), r in ROLES) # maintain UR
58         USERS.remove(user)
59
60      def AddRole(role):              # pre: role not in ROLES
61         ROLES.add(role)
62
63      def DeleteRole(role):           # pre: role in ROLES
64         UR -= setof((u,role), u in USERS) # maintain UR
65         PR -= setof((p,role), p in PERMS) # maintain PR
66         ROLES.remove(role)
67
68      def AddPerm(perm):              # pre: perm not in PERMS
69         PERMS.add(perm)
70
71      def DeletePerm(perm):           # pre: perm in PERMS
72         PR -= setof((perm,r), r in ROLES) # maintain PR
73         PERMS.remove(perm)
74
75      def AddUR(user, role):
76         # pre: user in USERS, role in ROLES, (user,role) not in UR
77         UR.add((user,role))
78
79      def DeleteUR(user, role):       # pre: (user,role) in UR
80         UR.remove((user,role))
81
82      def AddPR(perm, role):
83         # pre: perm in PERMS, role in ROLES, (perm,role) not in PR
84         PR.add((perm,role))
85
86      def DeletePR(perm, role):       # pre: (perm,role) in PR
87         PR.remove((perm,role))
88
89      def AssignedUsers(role):        # pre: role in ROLES
90         return setof(u, (u,_role) in UR)
91
92      def AssignedRoles(user):        # pre: user in USERS
93         return setof(r, (_user,r) in UR)
94
95      def UserPermissions(user):      # pre: user in USERS
96         return setof(p, (_user,r) in UR, (p,r) in PR)
97
98      def CheckAccess(user, perm):    # pre: user in USERS, perm in PPRMS
99         return some(r in ROLES, has= (user,r) in UR and (perm,r) in PR)
100
```

```
101
102  class HierarchicalRBAC_set(CoreRBAC,process):    # using while for Trans
103
104    def Trans(E):
105      T = E
106      while some((x,y) in T, (y,z) in E, has= (x,z) not in T):
107        T.add((x,z))
108      return T | setof((r,r), r in ROLES)
109
110  class HierarchicalRBAC_rules(CoreRBAC,process): # using rules for Trans
111
112    def rules(name= 'Trans_rules'):
113      if edge(x,y): path(x,y)
114      if edge(x,z) and path(z,y): path(x,y)
115
116    def Trans(E):
117      return infer(path, edge=E, rules=Trans_rules) | setof((r,r), r in ROLES)
118
119  class HierarchicalRBAC(HierarchicalRBAC_set,process):
120    """
121    Hierarchical RBAC keeps also a role hierarchy:
122
123      RH: set of pairs of roles, called ascendant and descendant roles,
124      where an ascendant role inherits permissions from a descendant role
125
126    with constraints:
127
128      RH subset ROLES * ROLES, and RH is acyclic
129
130    update functions for RH, subject to the constraints above:
131
132      AddInheritance(asc, desc)
133      DeleteInheritance(asc, desc)
134      with the same kinds of pre-conditions as updates in CoreRBAC
135
136    query functions including the following:
137
138      Trans:
139        the transitive closure of role hierarchy union reflexive role pairs
140      AuthorizedUsers(role):
141        the set of users of role or ascendant roles of role
142      AuthorizedRoles(user):
143        the set of roles of user or descendant roles of the roles
144    """
145
146    def setup():
147      self.RH = set()             # RH subset ROLES * ROLES, where asc inh desc
148
149    def AddInheritance(a, d):
150      # pre: a in ROLES,d in ROLES, (a,d) notin RH, a!=d, (d,a) notin Trans(RH)
151      RH.add((a,d))
152
153    def DeleteInheritance(a, d): # pre: (a,d) in RH
154      RH.remove((a,d))
155
156    def AuthorizedUsers(role):
157      return setof(u, (u,asc) in UR, (asc,_role) in Trans(RH))
158
159    def AuthorizedRoles(user):
160      return setof(r, (_user,asc) in UR, (asc,r) in Trans(RH))
161
162
163  class CoreRBACwithSSD(CoreRBAC,process):
164    """
165    Core RBAC with SSD keeps also a set of SSD items, where each item has:
```

```
        a name,
        a set of roles, and
        a cardinality

    with constraints:

        all roles in all SSD items  subset ROLES
        for each SSD item, its cardinality is > 0 and < the number of its roles
        for each user, for each SSD item,
           the number of assigned roles (AssignedRoles) of the user
           that are in the item's set of roles is at most the item's cardinality

    update functions, subject to the constraints above:

        CreateSsdSet(name, roles, c): add SSD item having name, roles, c
        DeleteSsdSet(name): delete SSD item having name
        AddSsdRoleMember(name, role): add role to roles of SSD item having name
        DeleteSsdRoleMember(name, role): del role fr roles of SSD item having name
        SetSsdSetCardinality(name, c): set c to be card. of SSD item having name
        with the same kinds of pre-conditions as updates in CoreRBAC, except that
        all updates have also pre-conditions that no constraints will be violated

    query functions including the following:

        SsdRoleSets(): the set of names of SSD items
        SsdRoleSetRoles(name): the set of roles in SSD item having name
        SsdRoleSetCardinality(name): the cardinality of SSD item having name
    """

    def setup():
      self.SsdNAMES = set()   # set of names of constraints
      self.SsdNR = set()      # set of pairs of name and role
                              # SsdNR subset SsdNAMES * ROLES
      self.SsdNC = set()      # set of pairs of name and cardinality
                              # SsdNC: SsdNAMES -> int

    # constraint named SSD, as post condition for all updates
    def constraint(name='SSD'):
      return each(u in USERS, (name,c) in SsdNC, has=
                  countof(r, r in AssignedRoles(u), (_name,r) in SsdNR) <= c)

    def CreateSsdSet(name, roles, c):
      # pre: name not in SsdNAMES, roles subset ROLES, 1 <= c < count(roles)
      SsdNAMES.add(name)
      SsdNR |= setof((name,r), r in roles)
      SsdNC.add((name,c))

    def DeleteSsdSet(name):  # pre: name in SsdNAMES  #don't need post SSD
      SsdNR -= setof((name,r), r in SsdRoleSetRoles(name))
      SsdNC.remove((name,SsdRoleSetCardinality(name)))
      SsdNAMES.remove(name)          # delete ssd name last

    def AddSsdRoleMember(name, role):
      # pre: name in SsdNAMES, role in ROLES
      # pre: role not in SsdRoleSetRoles(name)
      SsdNR.add((name,role))

    def DeleteSsdRoleMember(name, role):
      # pre: name in SsdNAMES, role in SsdRoleSetRoles(name)
      # pre: c < SsdRoleSetCardinality(name)-1
      SsdNR.remove((name,role))

    def SetSsdSetCardinality(name, c):
      # pre: name in SsdNAMES, SsdRoleSetCardinality(name) != c
```

```
     SsdNC.remove((name,SsdRoleSetCardinality(name)))
     SsdNC.add((name,c))

  def SsdRoleSets():
    return SsdNAMES

  def SsdRoleSetRoles(name):        # pre: name in SsdNAMES
    return setof(r, (_name,r) in SsdNR)

  def SsdRoleSetCardinality(name):  # pre: name in SsdNAMES
    return anyof(c, (_name,c) in SsdNC)


class HierarchicalRBACwithSSD(HierarchicalRBAC,CoreRBACwithSSD,process):
  """
  Hierarchical RBAC with SSD combines all from
  Hierarchical RBAC and Core RBAC with SSD, except that
  the SSD constraint uses AuthorizedRoles in place of AssignedRoles.
  """

  def constraint (name= 'SSD'):
    return each(u in USERS, (name,c) in SsdNC, has=
                countof(r, r in AuthorizedRoles(u), (_name,r) in SsdNR) <=c)


class AdminRBAC(HierarchicalRBACwithSSD):
  """
  Administrative RBAC for HierarchicalRBACwithSSD
  has optimization and planning functions:

    MineMinRoles:
      find a smallest set of roles with UR' and PR' assignments
      such that UR' * PR' = UR * PR

    MineMinRoleAssignments:
      find a smallest set of UR' and PR' assignments
      such that UR' * PR' = UR * PR = UP

    GetRolesPlan(user, roles, acts):
      find a sequence of actions, i.e., updates, in acts that
      allows user to get roles

    GetRolesShortestPlan(user, roles, acts):
      find a shortest sequence of actions, i.e., updates, in acts that
      allows user to get roles

  Any subset of updates can be used as acts.
  All constraints must hold after each action.

  The first two can have a version that includes finding RH'.

  Administrative RBAC could also be for
    CoreRBAC, HierarchicalRBAC, or CoreRBACwithSSD.
  """

  def MineMinRoles():
    return anyof((R, UR2, PR2), R in subset(ran(UR)&ran(PR)),
                  UR2 in subset(dom(UR)*R), PR2 in subset(dom(PR)*R),
                  UR2 * PR2 == UR * PR, minimize= count(R))

  def MineMinRoleAssignments():
    return anyof((R, UR2, PR2), R in subset(ran(UR)&ran(PR)),
                  UR2 in subset(dom(UR)*R), PR2 in subset(dom(PR)*R),
                  UR2 * PR2 == UR * PR, minimize= count(UR2+PR2))
```

```
296    def GetRolesPlan(user, roles, acts):
297      assume(True)
298      seq = []
299      while not each(r in roles, has= (_user,r) in UR):
300        if some(a in instances(acts)):
301          do(a)
302          seq.append(a)
303      achieve(anyof(seq))
304
305    def GetRolesShortestPlan(user, roles, acts):
306      assume(True)
307      seq = []
308      cost = 0
309      while not each(r in roles, has= (_user,r) in UR):
310        if some(a in instances(acts)):
311          do(a)
312          seq.append(a)
313          cost += 1
314      achieve(anyof((seq, cost), minimize= cost))
315
316
317  class DistRBAC(HierarchicalRBACwithSSD,process):
318    """
319    A Distributed RBAC process keeps also the following sets:
320
321      OTHERS: set of other RBAC processes
322      GuestR: set of pairs of a rbac-role pair and a guest role
323
324    with constraints:
325
326      domain(domain(GuestR)) subset OTHERS
327      range(GuestR) subset ROLES
328
329    update functions for each set subject to the constraints above:
330
331      AddGuestRole, DeleteGuestRole
332      AssignGuestRole:
333        assign to user of role in rbac the corresponding guest roles
334      DeassignGuestRole
335        deassign from user of role in rbac the corresponding guest roles
336
337    query functions:
338
339      GuestRoles (rbac,role): the set of guest roles for role of rbac
340      OthersRoles(guest): the set of rbac-role pairs for role guest
341
342    Distributed RBAC can also be for only
343      CoreRBAC, HierarchicalRBAC, or CoreRBACwithSSD,
344      or Administrative RBAC for any of these.
345    """
346
347    def setup(OTHERS):
348      self.GuestR = set()
349
350    def AddGuestRole(rbac, role, guest):     # pre: rbac in OTHERS,guest in ROLES
351      GuestR.add(((rbac,role),guest))
352
353    def DeleteGuestRole(rbac, role, guest):  # pre: ((rbac,role),guest) in GuestR
354      GuestR.remove(((rbac,role),guest))
355
356    def GuestRoles(rbac, role):
357      return setof(guest, ((_rbac,_role),guest) in GuestR)
358
359    def OthersRoles(guest):
360      return setof((rbac,role), ((rbac,role),_guest) in GuestR)
```

```
def AddGuestUR(user, rbac, role):          # pre: rbac in OTHERS
  send(('credential', user, role), to= rbac)
  if await(received(('accept', user, role), from_= rbac)):
    for r in GuestRoles(rbac, role):
      AddUR(user, r)

def DeleteGuestUR(user, rbac, role):
  for r in GuestRoles(rbac, role):
    DeleteUR(user, r)

def receive(msg=('credential', user, role), from_= rbac):
  if (user,role) in UR:
    send(('accept', user, role), to= rbac)
  else:
    send(('reject', user, role), to= rbac)

def receive(msg=('AddGuestUR', user, rbac, role)):
  AddGuestUR(user, rbac, role)

def receive(msg=('DeleteGuestUR', user, rbac, role)):
  DeleteGuestUR(user, rbac, role)
```

# Appendix I

# Security Policies in Constraint Handling Rules

Position Paper

Thom Frühwirth
Ulm University, Germany
http://www.informatik.uni-ulm.de/pm/fileadmin/pm/home/fruehwirth/

**Abstract**

Role-Based Access Control (RBAC) is a popular security policy framework. User access to resources is controlled by roles and privileges. Constraint Handling Rules (CHR) is a rule-based constraint logic language. CHR has been used to implement security policies for trust management systems like RBAC for more than two decades.

Constraint Handling Rules (CHR) [Frü09, Frü18, Frü15, FR18] is a logical rule-based language and framework employing constraints. In this short paper, we describe work on RBAC that is implemented in CHR. We just cite the main works for each group of authors. Further references may be found in the cited papers and/or by googling.

Based on [BFM02], Bistarelli et. al. [BMS10, BMS12] apply an extension of Datalog by weighted facts to model role-based trust management. Deduction can validate access requests. Abduction can compute missing credentials if the access is denied and it can compute the level of preference that would grant the access. Both deduction and abduction are expressed in Weighted Datalog and translated into CHR for execution. [BCMS14] show how this deductive and abductive reasoning can be efficiently ported to Android enabling distributed authorization. Both deduction and abduction are implemented as programs in a version of CHR that is embedded into Java (JCHR).

Ribeiro et. al. [RG99] present a static analyzer that automatically verifies consistency of workflow specifications written in WPDL (Workflow Process Definition Language) and of specifications in a security policy language (SPL). The analyzer is implemented with CHR embedded in SICstus Prolog. [RZFG00] further describes this Policy Consistency Verifier (PVC). It now includes constraints automatically annotated with temporal information. [RF07] presents further work on the security policy language (SPL). It can express the concepts of permission and prohibition, and some restricted forms of obligation as well as history-based approaches. Given a SPL specification, it is verified using CHR and then compiled to Java into a corresponding security access monitor. The current CHR verifier has about 300 rules and is able to solve all SPL constraints, including the constraints implicitly qualified with time.

The Object Constraint Language (OCL) is a declarative text language describing rules applying to Unified Modeling Language (UML) models. OCL provides constraint and object query expressions on models that cannot be expressed by diagrammatic notation. OCL is now a key component of the new OMG standard recommendation for transforming models. Model finders automatically verify UML/OCL models by checking satisfiability (consistency) of models using example instances. The work of [DTVH16] presents oclCHR `https://uet.vnu.edu.vn/~hanhdd/oclCHR/`, a verifier implemented in CHR embedded in Eclipse Prolog. It is of interest here, because the authors use an UML model of RBAC as their main example.

Finally, we would like to cite two approaches of RBAC in logical languages that can be readily translated into CHR. [BS03] show how a range of role-based access control (RBAC) models may be usefully represented as executable logical specifications in constraint logic programs (CLP). Like Weighted Datalog, CLP clauses can be translated to CHR [Frü09].

[OPR18] presents a declarative interpretation of Access Permissions (APs) as Linear Concurrent Constraint Programs (LCC). By interpreting LCC programs as formulae in intuitionistic linear logic, they can verify properties of AP programs such as deadlock detection, correctness, and concurrency. CHR also admits a linear logic interpretation [Bet14] and is closely related to the more recent LCC language. Translations between LCC and CHR are given in [Mar10].

Concluding, CHR is a often used language to build reasoning services. In this paper, we showed this surveying shortly work on the problem of security policies, i.e. access control. We would like to thank the anonymous referees for their helpful comments.

Appendix J

# How to upgrade ASP for true dynamic modelling and solving?

Torsten Schaub*

Inria, Rennes, France  and  University of Potsdam, Germany

`torsten@cs.uni-potsdam.de`

## 1  Motivation

*The world is dynamic, and ASP is not!* This is a provocative way to say that ASP is not up to dealing with many complex real-world applications having a dynamic nature, let alone transitions over states, not even mentioning more fine-grained temporal structures.

Although ASP has already been applied in various domains in academia and industry with remarkable success [8, 9], a closer look reveals that this concerns mostly static or smaller dynamic domains. For example, ASP is highly competitive in static domains such as timetabling [2] and workforce management [19], whereas it lags behind when it comes to substantial dynamic ones, as for instance robotic intra-logistics as discussed below. In fact, there is still quite a chasm between its level of development for addressing static and dynamic domains. This is because its modeling language as well as its grounding and solving machinery aims almost exclusively at handling static knowledge, while dynamic knowledge is usually indirectly dealt with via reductions to the static case.

In order to overcome this barrier and to upgrade ASP to the next level, dealing with complex dynamic problems, three areas appear to be relevant to me.[1]

## 2  Modeling

The most popular languages for modeling dynamic systems in the realm of ASP are temporal extensions of Equilibrium Logic [1], the host logic of ASP, and action languages [14]. Although both constitute the main directions of non-monotonic temporal systems, their prevalence lags way behind the usage of plain ASP for

---

[1]This personal view comes with a lot of self references — sorry for that!

modeling dynamic domains. Hence, notwithstanding the meticulous modeling of dynamics in ASP due to an explicit representation of time points, it seems that its pragmatic advantages, such as its rich (static) modeling language and readily available solvers, often seem to outweigh the firm logical foundations of both dedicated approaches. Although the true reasons are arguably inscrutable, let us discuss some possible causes.

The appeal of action languages lies in their elegant syntactic and semantic simplicity: they usually consist of static and dynamic laws inducing a unique transition system. Although most of them are implemented in ASP, their simplicity denies the expressive modeling possibilities of ASP. Also, despite some recent reconciliation [16], existing action languages lack the universality of ASP as reflected by the variety of variants.

Temporal Equilibrium Logic (TEL; [1]) builds upon an extension of the logic of Here-and-There [15] with Linear Temporal Logic (LTL; [18]). This results in an expressive non-monotonic modal logic, which relies upon the general syntax of LTL and possesses a computational complexity beyond LTL [3]. As in LTL, a model in TEL is an *infinite* sequence of states, called a trace. This rules out computation by ASP technology (and necessitates model checking) and is somewhat unnatural for applications like planning, where plans amount to finite prefixes of one or more (infinite) traces.

One proposal to overcome these issues is to restrict TEL to finite traces, similar to the restriction of LTL to $LTL_f$ advocated in [5]. This is detailed in [4] and accompanied with an extension of the ASP system *clingo*, dubbed *telingo* and available at https://github.com/potassco/telingo. *telingo* extends the full-fledged input language of *clingo* with temporal operators and computes temporal models incrementally by multi-shot solving [11].

## 3   Encoding and Solving

The need for dedicated encoding and solving techniques for handling dynamic domains stems from the necessity to implement fluents, that is, propositions changing their value over time. In ASP, just as other constraint-based approaches like CP or SAT, this amounts to creating a copy of each fluent and related rules per time point. The reduction of the resulting redundancy is the primary target of the aforementioned dedicated reasoning techniques.

First of all, we should realize that modeling and encoding a dynamic domain may amount to quite different specifications, both being declarative but aiming at different conceptions at distinct levels of the domain.[2] The easiest way to realize the difference between modeling and encoding is to consider a temporal rule $a(X) \leftarrow \bullet b(X)$ in which '$\bullet$' denotes the "previous" operator, or in *telingo* syntax: `a(X) :- 'b(X)`, that is finally encoded as `a(X,t) :- b(X,t-1)` where

---

[2]Another good example for this are (arithmetic) CSPs, nicely modeled by expressions like $x + y < 7$ but usually best implemented in ASP (and SAT) via an order encoding [23] treating integer variables by inequalities like, $x \leq 1, x \leq 2, \ldots$ (rather than a direct encoding using equalities $x = 1, x = 2, \ldots$).

the parameter 't' is handled by multi-shot solving [11]. Obviously, modeling is ideally more abstract than encoding by dropping aspects like the implementation of time by increasing integers. Also, the targeted implementation using parameters 't' (instead of variables 'T') remains hidden. But apart from this abstraction, no real gain is obtained as regards the elimination of redundancy. Unlike this, multi-shot solving cuts back redundancies by avoiding repeated grounding and solving efforts.

Much more is possible. Encoding-wise an exemplar is the *parallel* representation of sequential plans which has been investigated in SAT planning [7, 21]. A first attempt to transfer this to ASP is given in [6].[3] Another example is multi-path planning in logistics warehouses, where a two-step abstraction encoding technique was used [17] to scale up to state of the art algorithms. Certainly, many more such principled techniques exist but are no matter of common knowledge. Solving-wise, the semantic links between the aforementioned fluent copies need to be exploited. Again, SAT planning serves us as an exemplar, where heuristics were used in [20] to provide such links. This idea has led to the heuristic directives in *clingo* [12]. Another solving technique was put forward in [10], where ground multi-state invariants are extracted and generalized in order to be fed back into the solving process, thus extending their scope to all similar state combinations.[4] And much more can and needs to be done!

## 4   Benchmarking

The upgrade of ASP is moreover threatened by a lack of complex benchmark scenarios mimicking the needs of dynamic real-world applications. In contrast to many available benchmark suites, often supplied by automatic instance generators, real-world applications are rarely disseminated, either because they are classified or come only with a handful of instances. Another commonality of existing benchmark suites is that they are kept simple, stick to basic ASP, and usually feature at most a single specifics, so that they can be processed by as many systems as possible. However, this is in contrast to many real-world applications whose solution requires the integration of multiple types of knowledge and forms of reasoning. Last but not least, a feature distinguishing ASP from all other solving paradigms is its versatility, which is best put in perspective by solving multi-faceted problems.

The fear is thus that the lack of complex benchmark scenarios becomes a major bottleneck in ASP's progression towards real-world applications, and hence that more and more should be made available to our community. As a first step to overcome this problem, we propose in [13] the domain of *robotic intra-logistics*, a key domain in the context of the fourth industrial revolution, as witnessed by Amazon's Kiva, GreyOrange's Butler, and Swisslog's CarryPick systems.[5]

---

[3] This is implemented in the *plasp* system available at https://github.com/potassco/plasp.
[4] This is implemented in the *ginkgo* system available at https://github.com/potassco/ginkgo.
[5] www.amazonrobotics.com, www.greyorange.com/products/butler, www.swisslog.com/

All of them aim at automating warehouse operations by using robot vehicles that drive underneath mobile shelves and deliver them to picking stations. From there, workers pick and place the requested items in shipping boxes. Apart from the great significance of this real-world domain, our choice is motivated by several aspects. First of all, the domain is highly dynamic. At the same time, the warehouse layout is grid-based and thus provides a suitable abstraction for modeling robot movements in ASP. Moreover, the domain offers a great variety of manifold problem scenarios that can be put together in an increasingly complex way. For instance, one may start with single or multi-robot path-finding scenarios induced by a set of orders that are accomplished by using robots for transporting shelves to picking stations. This can be extended in various ways, for example, by adding shelf handling and delivery actions, considering order lines with multiple product items, keeping track of the number of ordered and/or stored product items, modeling energy consumption and charging actions, taking into account order frequencies, supplies, and priorities, striving for effective layouts featuring dedicated locations, like highways or storage areas, and so on. Finally, the domain is extremely well-suited for producing scalable benchmarks by varying layouts, robots, shelves, orders, product items, etc. Inspired by this, we have developed the benchmark environment *asprilo* [13] consisting of four parts (i) a benchmark generator, (ii) a solution checker, (iii) a benchmark and solution visualizer, and (iv) a variety of reference encodings. The design of *asprilo* was driven by the desire to create an easily configurable and extensible framework that allows for generating scalable, standardized benchmark suites that can be visualized with and without a corresponding solution. Correctness can be established via a modular solution checker. The accompanying reference encodings may serve as exemplary bases for extended encodings addressing more complex scenarios. The *asprilo* framework is freely available at `https://potassco.org/asprilo`.

## 5 Summary

Many people well beyond our community get interested by the modeling and solving capabilities of ASP, its elegance, succinctness, transparency, and last but not least its effectiveness. We attract them by showcasing *our* exemplary problems and solutions. But at the end of the day, if we want to keep them interested in ASP, we have to solve *their* problems.

---

`carrypick`

# Appendix K
# A Picat-based XCSP Solver – from Parsing, Modeling, to SAT Encoding

Neng-Fa Zhou[1] and Håkan Kjellerstrand[2]

[1] CUNY Brooklyn College & Graduate Center
[2] hakank.org

**Abstract.** This short paper gives an overview of a Picat-based XCSP3 solver, named PicatSAT, submitted to the 2018 XCSP competition. The solver demonstrates the strengths of Picat, a logic-based language, in parsing, modeling, and encoding constraints into SAT.

**XCSP3**

XCSP3 [1] is an XML-based domain specific language for describing constraint satisfaction and optimization problems (CSP). XCSP3 is positioned as an intermediate language for CSPs. It does not provide high-level constructs as seen in modeling languages. However, XCSP3 is significantly more complex than a canonical-form language, like FlatZinc. A constraint can sometimes be described in either the standard format or simplified format. The advanced format, which is used by group and matrix constraints, allows more compact description of constraints.

**Picat**

Picat [6] is a simple, and yet powerful, logic-based multi-paradigm programming language. Picat is a Prolog-like rule-based language, in which predicates, functions, and actors are defined with pattern-matching rules. Picat incorporates many declarative language features for better productivity of software development, including explicit non-determinism, explicit unification, functions, list comprehensions, constraints, and tabling. Picat also provides imperative language constructs, such as assignments and loops, for programming everyday things. Picat provides facilities for solving combinatorial search problems, including a common interface with CP, SAT, and MIP solvers, tabling for dynamic programming, and a module for planning. PicatSAT uses the SAT module, which generally performs better than the CP and MIP modules on competition benchmarks.

**Parsing**

The availability of different formats in XCSP3 makes it a challenge to parse the XCSP3 language. A parser implemented in C++ by the XCSP designers has

more than 10,000 lines of code. The entire Picat implementation of XCSP3 has about 2,000 lines of code, two-thirds of which is devoted to parsing and syntax-directed translation. As illustrated in the following example, Picat is well suited to parsing.

```
% E -> T E'
ex(Si,So) => term(Si,S1), ex_prime(S1,So).

% E' -> + T E' | - T E' | e
ex_prime(['+'|Si],So) =>
    term(Si,S1),
    ex_prime(S1,So).
ex_prime(['-'|Si],So) =>
    term(Si,S1),
    ex_prime(S1,So).
ex_prime(Si,So) => So = Si.
```

The parser follows the framework for translating context-free grammar into Prolog [3]: a non-terminal is encoded as a predicate that takes an input string (`Si`) and an output string (`So`), and when the predicate succeeds, the difference `Si-So` constitutes a string that matches the nonterminal. Unlike in Prolog, pattern-matching rules in Picat are fully indexed, which facilitates selecting right rules based on look-ahead tokens.

### Modeling

It is well known that loops and list comprehensions are a necessity for modeling CSPs. The following Picat example illustrates the convenience of these language constructs for modeling.

```
post_constr(allDifferentMatrix(Matrix)) =>
    NRows = len(Matrix),
    NCols = len(Matrix[1]),
    foreach (I in 1..NRows)
        all_different(Matrix[I])
    end,
    foreach (J in 1..NCols)
        all_different([Matrix[I,J] : I in 1..NRows])
    end.
```

The `allDifferentMatrix(Matrix)` constraint takes a matrix that is represented as a two-dimensional array, and posts an `all_different` constraint for each row and each column of the matrix.

### SAT Encoding

PicatSAT adopts the log encoding for domain variables. While log encoding had been perceived to be unsuited to arithmetic constraints due to its hindrance

to propagation [2], we have shown that log encoding can be made competitive with optimizations [4]. There are hundreds of optimizations implemented in PicatSAT, and they are described easily as pattern-matching rules in Picat. We have also shown that, with specialization, the binary adder encoding is not only compact, but also generally more efficient than BDD encodings for PB constraints [5]. PicatSAT adopts specialized decomposition algorithms for some of the global constraints. While competitive overall, PicatSAT is not competitive with state-of-the-art CP solvers on benchmarks that use path-finding constraints that require reachability checking during search. The future work is to design efficient encodings for these global constraints.

# Appendix L

# A software system should be declarative except where it interacts with the real world

Peter Van Roy*
Université catholique de Louvain

May 30, 2018

## Abstract

We propose a system design principle that explains how to use declarative programming (logic and functional) together with imperative programming. The advantages of declarative programming are well known; they include ease of analysis, verification, testing, optimization, maintenance, upgrading, and distributed implementation. We do not elaborate on the advantages here, but rather focus on what part of the software system should be written declaratively. We observe that declarative programming cannot interact directly with the real world while remaining declarative, since it does not support common real-world concepts such as physical time and named state. It is important to distinguish reasoning about the real world from interacting with the real world: declarative programming can do the first but not the second. Other programming paradigms that support these concepts must be used, such as imperative programming (which contains named state). To optimize the system design, we propose that real-world concepts should only be used where they cannot be avoided, namely where the system interfaces with the real world. It follows that a software system should be built completely declaratively except possibly where it interfaces with the real world. We motivate this principle with examples from our research and we outline a formal argument to justify it.

## 1 Introduction

The interplay between declarative (e.g., pure functional or logic) programming and imperative programming (which uses named mutable state) has long been a subject of debate in software design. Important questions are which paradigm to use and when to use it; and when and how to use the paradigms together. The book [1] presents these paradigms in a uniform framework, each with its kernel language, and carefully explains what each can and cannot do. This shows that there is no one paradigm that is uniformly better than the others. Large programs will typically use different paradigms in different parts, just as building a house requires multiple skills such as masonry, carpentry, plumbing, and electricity. But determining which paradigm to use where is left unanswered. This position paper gives a design principle that answers this question:

> A software system should be built completely with declarative programming except where it interacts with the real world.

Section 2 gives two examples to motivate this principle. Section 3 defines what we mean by *interaction with the real world* and gives a formal argument to support the principle. Section 4 discusses some ramifications of the principle, and Section 5 presents a brief conclusion.

## 2 Motivation

### 2.1 Example 1: client/server

Consider a client/server application in its simplest form: two clients communicating with one server. Each client sends requests to the server and receives replies. To satisfy liveness of each client, the server must accept each incoming request and reply to it within a reasonable delay. The server's handling of a client request should not be impeded because of what the other client does or does not do. The order of the server's handling of requests cannot be determined in advance, because it depends on the precise timing of the requests.

### 2.2 Example 2: convergent computation (CRDTs and Lasp)

We give a more substantial example taken from our research into synchronization-free programming for distributed systems, namely the Lasp programming system [2]. A Lasp program consists of a dataflow graph connecting data structures with functional and logical operations (similar to SQL operations). The data structures and operations are both designed to do *convergent computation*: each computation step adds information monotonically. In fact, the data structures are CRDTs (Conflict-free Replicated Data Type) [3]. To the programmer, Lasp executes as a functional/logic program with a dataflow semantics, similar to the deterministic dataflow of Chapter 4 in [1].

The data structures are replicated and information is periodically disseminated between replicas. CRDTs are designed so that the replicas are always converging to the correct result. This is extremely resilient. Changing the timing of the dissemination messages has no effect on correctness. Dropped, delayed, or reordered messages have no effect on correctness. The only possible effect is to slow down convergence. Node crashes have two possible results: either the crash has no effect except for slowing down convergence, or some information disappears completely.

Lasp provides a functional semantics with a highly resilient distributed implementation based on weak synchronization. A Lasp program needs stronger synchronization than periodic dissemination only when talking to external clients (i.e., real-world interaction).

## 3 Formal definition

We have given two examples that show the principle in action. We now outline a formal argument to justify the principle and to define it precisely. For brevity, we use the lambda calculus, but any other declarative execution model that performs deduction from an initial input could be used instead.

### 3.1 Functional programming

Consider the lambda calculus with a concurrent evaluation strategy. Given an initial lambda expression, we can reduce it to its normal form. Any reduction order will lead to the same normal form, which is known as the Church-Rosser property. The reduction may take many steps. At any step in the reduction, there may be several different positions in the lambda expression that can be reduced next. By adding a scheduler to determine how to choose the position reduced at each step, we can define a concurrent form of functional programming very similar to the deterministic dataflow model given in Chapter 4 of [1]. In this form, we can define deterministic concurrent agents that communicate through streams, similar to Kahn networks [4].

### 3.2 Real-world interaction requires extending this model with time

Real-world interaction requires that the computation take into account input coming from the real world. If there are several inputs, they need to be handled in some order. This order is not known in advance; it becomes known during the reduction. We are led to a sequence of inputs, arriving one by one during the reduction process.

If all inputs would be known in advance, then they could be considered part of the initial expression, and the system would be purely declarative. However, a key property of the real world is that they are not known in advance. They arrive during the reduction process because reduction steps take nonzero time. The arrival order is determined by the precise timing with respect to the reduction process. The order can affect the result. For example, if the computation builds a list, the order of its elements can depend on the timing. We conclude that the new concept that must be added to deterministic dataflow to allow interaction with the real world is *time*.

It may be that an expression is not reducible until an input arrives, in which case we say the reduction is suspended. When the input arrives, a reduction step becomes possible. When this step is taken, we say the reduction *synchronizes* with the input. On the other hand, the expression may be reducible at several positions, and an arriving input creates another position where a reduction is possible. In this case, the system is active but can accept new input during execution.

## 4 Discussion

We claim that the design principle holds generally whenever a system interacts with the real world. The real-world property "time" can appear in different guises to the system, e.g., as nondeterminism (see above), as physical wall-clock time (hardware clock), or as partial failure.

Interaction with the real world happens not just at the API, but everywhere that the real world has to be taken into account. For example, MapReduce handles a straggler (slow node) by speculatively running a copy of its task on another node. Detecting the slowness of a straggler depends on time, which is an interaction with the real world.

## 5 Conclusions

This position paper presents a software design principle that is a result of the author's study of the differences between declarative and imperative programming for system building. We are working on a full paper with a detailed formal justification of the principle and its application to synchronization-free programming.


\end{appendices}

\end{document}